\documentclass[conference,compsoc]{IEEEtran}

\usepackage{amsmath,amssymb}

\usepackage[nocompress]{cite}
\usepackage[hyphens]{url}
\usepackage[hidelinks]{hyperref}

\usepackage[utf8]{inputenc}
\usepackage[english]{babel}

\usepackage{graphicx}
\DeclareGraphicsExtensions{.pdf,.png,.jpg}

\usepackage{booktabs}
\usepackage{tabularx}
\usepackage{multirow}
\usepackage{enumitem}
\usepackage{makecell}
\usepackage{relsize}
\usepackage{placeins}

\usepackage{xcolor}
\usepackage{colortbl}
\usepackage[most]{tcolorbox}
\usepackage[normalem]{ulem}
\usepackage{tikz}
\usetikzlibrary{arrows.meta,positioning,shapes,calc}

\definecolor{darkgreen}{RGB}{0,100,0}
\definecolor{codegreen}{rgb}{0,0.5,0}
\definecolor{codepurple}{rgb}{0.58,0,0.82}
\definecolor{codegray}{rgb}{0.5,0.5,0.5}
\definecolor{cdnavy}{HTML}{17324D}
\definecolor{cdblue}{HTML}{2F6690}
\definecolor{cdteal}{HTML}{2A7F62}
\usepackage{listings}
\lstdefinestyle{mystyle}{
  commentstyle=\color{codegreen},
  keywordstyle=\bfseries,
  stringstyle=\color{codepurple},
  basicstyle=\ttfamily\scriptsize,
  breaklines=true,
  captionpos=b,
  keepspaces=true,
  tabsize=2
}
\lstdefinestyle{casecode}{
  style=mystyle,
  language=C,
  basicstyle=\ttfamily\fontsize{6.4pt}{7.3pt}\selectfont,
  numbers=left,
  numberstyle=\ttfamily\fontsize{5.2pt}{6pt}\selectfont\color{codegray},
  numbersep=4pt,
  frame=single,
  framerule=0.35pt,
  rulecolor=\color{black!30},
  backgroundcolor=\color{black!1},
  framesep=2pt,
  xleftmargin=2.8em,
  framexleftmargin=2.3em,
  xrightmargin=0.3em,
  framexrightmargin=0.3em,
  aboveskip=2pt,
  belowskip=4pt,
  breaklines=true,
  breakatwhitespace=false,
  columns=fullflexible,
  showstringspaces=false
}

\newtcblisting[auto counter]{concurcodelisting}[2][]{
  enhanced,
  float=t,
  listing only,
  colback=black!1,
  colframe=cdnavy,
  colbacktitle=cdnavy,
  coltitle=white,
  title={Listing~\thetcbcounter. #2},
  fonttitle=\bfseries\scriptsize,
  boxrule=0.6pt,
  arc=1mm,
  left=1mm,right=1mm,top=0.7mm,bottom=0.7mm,
  listing options={
    language=C,
    basicstyle=\ttfamily\scriptsize,
    commentstyle=\color{cdteal},
    keywordstyle=\bfseries\color{cdnavy},
    breaklines=true,
    keepspaces=true,
    columns=fullflexible
  },
  #1
}

\usepackage[noend,ruled,linesnumbered]{algorithm2e}
\IfFileExists{todonotes.sty}{%
  \usepackage{todonotes}
}{%
  \newcommand{\todo}[2][]{\textcolor{red}{\textbf{TODO:} #2}}
}

\newcommand{\find}[1]{%
\begin{tcolorbox}[tile,size=fbox,boxsep=2mm,boxrule=0pt,top=0pt,bottom=0pt,
borderline={0.6mm}{0pt}{black!66!white},colback=black!5!white]
\em #1
\end{tcolorbox}
}

\newcommand{\denseitems}{\setlength{\itemsep}{1pt}\setlength{\parskip}{0pt}}

\newif\ifcomments
\commentstrue
\ifcomments
  
  \newcommand{\wen}[1]{\todo[inline,color=red!40,linecolor=red,bordercolor=red]{Wen: #1}}
\else
  
  \newcommand{\wen}[1]{}
\fi

\definecolor{DarkOrange}{rgb}{0.8,0.3,0.0} 
\definecolor{DarkCyan}{rgb}{0.0, 0.55, 0.55}
\definecolor{codegreen}{rgb}{0,0.6,0}
\definecolor{codegray}{rgb}{0.5,0.5,0.5}
\definecolor{codepurple}{rgb}{0.58,0,0.82}
\definecolor{backcolour}{rgb}{0.95,0.95,0.92}

\newcommand{\tech}{\mbox{\textsc{ConcurDep}}}

\newcommand{\paperkeywords}{Concurrency, Program Analysis, CPython, Memory Safety, Free Threading}

\newcommand{\circleone}[1]{%
  \tikz[baseline=(char.base)]{
    \node[
      shape=circle,
      draw=black,
      fill=white,
      text=black,
      inner sep=1pt
    ] (char) {\small #1};
  }%
}

\newcommand{\ncdg}{\textsc{NCDG}}

\newif\ifcamera
\cameratrue % flip to \cameratrue for camera-ready

\title{\Large \bf \tech{}: Event-Guided Analysis of Dependency Invalidation in CPython Concurrency}

\ifcamera
  \author{
  \IEEEauthorblockN{Baihong Chen}
  \IEEEauthorblockA{Utah State University}
  \and
  \IEEEauthorblockN{Hadley Westover}
  \IEEEauthorblockA{Utah State University}
  \and
  \IEEEauthorblockN{Wen Li}
  \IEEEauthorblockA{Utah State University}
  }
\fi

\begin{document}
\bstctlcite{concurdep:BSTcontrol}
\maketitle

%!TEX root = paper.tex

\begin{abstract}
Removing CPython's Global Interpreter Lock (GIL) exposes native code to concurrency
absent from ordinary C types. Mutation or re-entry can revoke a borrowed
object, storage pointer, traversal state, or lease between acquisition and use
while its owner remains alive, causing native memory errors and runtime-state
corruption. Race analyses track conflicting accesses. Python/C lifecycle
analyses track individual object states. These reporting units leave implicit
owner--subject--storage relations disconnected from later uses under parallel
and re-entrant events.

We present \tech{}, a source-level static analysis of \emph{dependency
invalidation}. Its key insight is to represent the runtime property a native use
requires and ask which target-matched event can revoke it within the dependency's
live region. \tech{} recovers runtime-semantic dependencies, connects them to
events through an event-aware native concurrency dependency graph, and applies
property-specific state and protection transfers through a shared engine and
six mechanism plugins.
\tech{} correctly classifies all 180 matched semantic-conformance cases and
analyzes each of three production CPython releases with five-run medians of
22.13--27.66 seconds and 670--784\,MiB peak resident memory.
Source auditing confirms 1,094 of 4,273 unique production fingerprints
(25.60\% confirmation yield).
Removing derived relation recovery loses 25--44 represented roots per
release; removing cross-entry events loses 73--94.
The study identifies 144 distinct bugs across free-threaded and
conventional-GIL builds, including 95 previously unreported in public
sources. These results show that explicit dependency, event, and property
semantics expose consequential runtime failures across API boundaries
and execution modes at release scale.

\end{abstract}

\begin{IEEEkeywords}
\paperkeywords
\end{IEEEkeywords}

\section{Introduction}\label{sec:intro}

CPython is replacing implicit synchronization with explicit concurrency. Python
Enhancement Proposal (PEP)~703 introduced a build with an optional Global
Interpreter Lock (GIL), allowing Python threads to execute in
parallel~\cite{pep703}. The transition required biased reference counting,
deferred reclamation, per-object locking, and critical-section application
programming interfaces (APIs)~\cite{pep703}. Extensions and runtime paths that
relied on the GIL to protect object or global state now synchronize that state
explicitly~\cite{py-free-threading-extensions}; failures can corrupt interpreter
state, violate native memory safety, or stop the process.

The difficult failures extend beyond unguarded loads and stores. Native CPython
code acquires relationships on which it later relies: a list owns a borrowed
element, a bytearray backs a raw pointer, a dictionary table supports traversal,
or a memoryview maintains a buffer lease. Mutation, reference drops, resource
release, finalization, or Python re-entry can revoke these relationships between
acquisition and use while the owner remains alive. The problem is to determine
which runtime event can invalidate which native dependency before which use.

Existing techniques supply parts of this reasoning. Dynamic race and memory-error
detectors expose executed conflicts and manifested faults~\cite{tsan,asan};
static race analyses relate shared accesses to aliasing and
synchronization~\cite{goblint,locksmith}. Python/C lifecycle analyses track
reference counts and object-state transitions~\cite{pungi,pyrefcon}, while
interference-aware typestate combines transitions with locksets
~\cite{typestate-concurrency}. CPython adds implicit owner--subject--storage
contracts and parallel or re-entrant entries. The analysis connects these
contracts to later uses and distinguishes ownership, leases, synchronization,
and revalidation. Such failures can span fields or allocations, beyond one
conflicting access or one object's lifetime transition.

This gap creates three design-independent challenges.
\emph{C1. dependency recovery} reconstructs implicit runtime relations and
use-time obligations from C operations, macros, and API contracts.
\emph{C2. event connection} determines whether an invalidating event can occur
within the dependency's live region, target its runtime state, and overlap through
a parallel or re-entrant context.
\emph{C3. property-specific validation} tracks the distinct properties an event
changes and the guarantees of locks, ownership, leases, and point checks.

\tech{} rests on one reasoning shape: an operation establishes an owner--subject
dependency $D$, an event $E$ may revoke a required property within $D$'s live
region, and a later use consumes that property. To address C1, \tech{} normalizes
native operations and recovers explicit owner--subject--storage dependencies.
Its event-aware native concurrency dependency graph (\ncdg{}) addresses C2
through interval, target, and overlap relations. A shared runtime-state engine
addresses C3 by applying target-matched effects and property-specific protection;
six mechanism plugins define use-time obligations. A declarative line-delimited
JSON (JSONL) catalog supplies versioned CPython API and type semantics.

We evaluate controlled and production behavior. \tech{} correctly classifies all
180 matched ConcurrBench cases; cumulative exposure, target/live-region, and
state/protection ablations each remove controls while retaining all positives.
Across five runs per CPython release, whole-program analysis takes
22.13--27.66 seconds at 670--784\,MiB median peak resident memory. Auditing every
unique frozen fingerprint on CPython 3.13--3.15 confirms 1,094 of 4,273 and
attributes 139 free-threaded root causes to those reports. Investigation found
two detector gaps; the released configuration recovers all 141 qualified
free-threaded roots in the study corpus. The cross-mode inventory contains 144
root causes, including 125 whose strongest evidence is native failure. A
public-history audit finds independent prior coverage for 49 roots and no
earlier public report of the same native root for 95. On production inputs,
one-capability-at-a-time ablations show that derived relation recovery retains
25--44 additional represented roots per release and cross-entry event connection
retains 73--94. A witness-bound audit changes neither semantic warnings nor
represented roots; it changes only alternative profile completeness.
SVF-MTA/SlicedMTA completes with zero race pairs, Goblint reaches the 24-hour
budget, RacerF and Deagle reject required features, and O2 is unavailable.
This paper makes three \textbf{contributions}:

\begin{itemize}[leftmargin=1.25em]
  \item We formulate native concurrency failures as property-specific
  invalidation of recovered owner--subject--storage dependencies, unifying
  parallel and re-entrant execution without reducing it to same-address conflicts.

  \item We develop an event-aware \ncdg{} and runtime-state analysis linking
  dependencies to interprocedural events by live region, target identity, and
  execution context, and validating use-time obligations under property-specific
  protection.

  \item We implement \tech{} and provide its analyzer, semantic catalogs,
  ConcurrBench, production reports, and auditable root-evidence ledger for three
  CPython releases. Evaluation establishes controlled semantic conformance,
  release-scale cost, and 144 qualified production root causes across the studied
  runtime configurations.
\end{itemize}

%\smallskip
%\noindent\textbf{Open science.}
%An \artifact{anonymized supplementary snapshot} accompanies the submission with
%the analyzer, semantic catalogs, controlled benchmark, raw reports, root-oriented
%bug packages, evidence ledger, and snapshot-build script. The builder records
%file identities. It strips repository history, remote URLs, hostnames, absolute
%workspace paths, and identifying metadata. Static candidates, qualification
%evidence, and disclosure status remain separate.

\section{Background and Motivation}\label{sec:background}

\subsection{Background}

Conventional CPython serializes most Python and C-API execution with the Global
Interpreter Lock (GIL), but callbacks, finalizers, signals, blocking operations,
and explicit GIL release still permit re-entry or overlap~\cite{py-gil-docs}.
Free-threaded builds allow parallel Python threads, relying instead on
thread-safe reference counting, per-object locks, critical sections, and delayed
reclamation~\cite{pep703}. These mechanisms protect particular operations and
properties, not arbitrary multi-operation invariants.
Native code often relies on runtime relationships absent from C types. We model
a relationship as a \emph{dependency}: an acquisition connects an owner to a
subject, and later uses require stable properties---object lifetime,
owner--subject membership, storage address, representation, traversal state,
buffer leases, or callback-sensitive state. This unit records why a native use
is valid, not just the address it accesses.

\subsection{Motivating Example}\label{sec:motiexample}

Consider two threads in a free-threaded CPython process sharing a
\texttt{memoryview} backed by a mutable \texttt{bytearray}.
Thread~A calls \texttt{memoryview.tobytes()}, while Thread~B releases
the same view and resizes the bytearray. The native copy relies on
valid access to the backing storage until completion.
Listing~\ref{fig:motivation} simplifies the native copy and the concurrent
Python operations. Here, \texttt{self} and \texttt{view} denote the same
memoryview, and \texttt{backing} is its bytearray exporter.

\begin{concurcodelisting}[label={fig:motivation},nofloat]{View release before a native backing-buffer use
in \texttt{memoryview.tobytes()}}
/* Thread A: memoryview.tobytes() */
src = VIEW_ADDR(self);          // a: get descriptor
CHECK_RELEASED(self);           // check current state
out = new_bytes(src->len);
copy_from_view(out, src);       // u: read backing

/* Thread B: simplified Python operations */
view.release();                 // e: release same view
backing.resize_and_refill();    // backing may move
\end{concurcodelisting}

\texttt{VIEW\_ADDR(self)} obtains a pointer to the view's private
\texttt{Py\_buffer} descriptor ($a$)~\cite{py-memoryview-api}.
The later copy ($u$), represented by \texttt{copy\_from\_view},
uses \texttt{PyBuffer\_ToContiguous} and relies on the view's existing
buffer export---its \emph{lease} on the backing
storage~\cite{py-buffer-protocol}.
If Thread~B calls \texttt{release()} ($e$) after the release check,
it can drop the managed buffer's export~\cite{py-memoryview-release}.
A subsequent resize may then replace or reclaim storage whose address
the copy still uses.

\noindent\textbf{Limitations and challenges.}
The example illustrates three analysis difficulties.
First, the type of \texttt{src} describes a buffer descriptor but does
not express the copy's dependence on the view's active export
(\textbf{C1}). Recovering that dependence requires following the
relationship among the view, its descriptor, and the exporter through
macros, fields, and API contracts.
Second, release changes the view's export state in another API
implementation, whereas the copy reads the exporter's data allocation
(\textbf{C2}). A conflict on the released flag alone does not explain
this connection. The analysis determines whether release can affect
the same view after the check and before the copy finishes; a release
of an unrelated view or one ordered after the copy does not explain
this failure.
Third, object lifetime, a successful state check, and an active export
provide different guarantees (\textbf{C3}). Keeping \texttt{self} alive
preserves the view object, while \texttt{CHECK\_RELEASED} observes only
its current state. Neither prevents later release. Other memoryview
paths record temporary exports in \texttt{self->exports} around native
buffer uses. Acquiring such an export atomically with the release check
and retaining it through the copy prevents this revocation.
The protected property is backing-address validity, not immutable
contents~\cite{py-thread-safety-guarantees}.

\noindent\textbf{Motivating our approach.}
The example motivates connecting what a native operation requires to
the events that can invalidate it. \tech{} recovers the copy's dependence
on the view's export, connects it to a possible release of that same
view during the copy, and checks whether the relevant storage remains
protected. These tasks motivate dependency recovery, event connection,
and property-specific validation, respectively.

\subsection{Threat Model}

We distinguish three classes of triggering behavior.
An \emph{ordinary-API} trigger shares mutable objects and invokes public
Python operations concurrently.
A \emph{callback-dependent} trigger additionally controls a documented
Python callback, finalizer, descriptor, or codec.
A \emph{native-harness} trigger uses a conforming C-API test callable
to exercise GIL-release or lifetime interactions within CPython's
extension model.
We target violations of CPython's native memory safety and documented
API behavior. C-API objects follow their contracts; exporters
that violate the buffer protocol are outside
scope~\cite{py-buffer-protocol}.
Static reports identify potential invalidations, while source audit
and dynamic or causal qualification establish concrete triggers and
observed consequences. The motivating example uses ordinary public
APIs; Section~\ref{sec:rq3} presents its qualification evidence.

\section{Design}\label{sec:tech}

%\subsection{Overview}\label{sec:overview}

Given native CPython source, a compilation database, and a runtime API catalog,
\tech{} identifies native uses whose required state a concurrent or re-entrant
event may invalidate. Reports retain the acquisition, later use, owner--subject
relation, possible invalidating event, protection evidence, and violated
stability property for auditing.

\begin{figure*}[t]
  \centering
  \begin{tikzpicture}[
      font=\sffamily,
      endpoint/.style={draw=cdnavy,line width=0.65pt,rounded corners=1mm,
                      minimum height=24mm,text width=16mm,align=center,
                      fill=cdnavy!3,inner sep=1.4mm,font=\scriptsize},
      stage/.style={draw=cdnavy,line width=0.65pt,rounded corners=1mm,
                    minimum height=24mm,text width=34mm,fill=white,
                    inner sep=0pt},
      header/.style={draw=none,fill=cdnavy!8,text=cdnavy,align=center,
                     minimum height=10.5mm,text width=32mm,inner sep=0.5mm,
                     font=\scriptsize\bfseries},
      body/.style={draw=none,align=left,text width=30mm,inner sep=0pt,
                   font=\fontsize{6.5}{7.2}\selectfont},
      flow/.style={-{Latex[length=1.8mm]},line width=0.75pt,draw=cdnavy},
      artifact/.style={font=\tiny\bfseries,text=cdnavy,fill=white,
                       inner sep=0.6mm}]

    \node[endpoint] (input) at (9.5mm,0) {
      {\bfseries\color{cdnavy}INPUTS}\\[1mm]
      source +\\semantic catalog};

    \node[stage] (recover) at (45mm,0) {};
    \node[header] at ($(recover.center)+(0,6.7mm)$) {
      Runtime-Semantic\\Dependency Recovery};
    \node[body] at ($(recover.center)+(0,-5mm)$) {
      $\bullet$ normalize native operations\\[0.6mm]
      $\bullet$ recover dependencies and contracts};

    \node[stage] (graph) at (84mm,0) {};
    \node[header] at ($(graph.center)+(0,6.7mm)$) {
      Event-Aware\\Native Concurrency\\Dependency Graph};
    \node[body] at ($(graph.center)+(0,-5mm)$) {
      $\bullet$ reuse calls and lift event effects\\[0.6mm]
      $\bullet$ match interval, target, and overlap};

    \node[stage] (check) at (123mm,0) {};
    \node[header] at ($(check.center)+(0,6.7mm)$) {
      Runtime-State-Aware\\Dependency Validation};
    \node[body] at ($(check.center)+(0,-5mm)$) {
      $\bullet$ propagate state and protection\\[0.6mm]
      $\bullet$ evaluate plugins and retain witnesses};

    \node[endpoint] (output) at (154.5mm,0) {
      {\bfseries\color{cdnavy}AUDIT REPORT}\\[1mm]
      witness\\[-0.2mm]$a\rightarrow e\rightarrow u$\\[0.7mm]
      violated property +\\missing protection};

    \draw[flow] (input.east) -- node[artifact,above] {$P{+}C$} (recover.west);
    \draw[flow] (recover.east) -- node[artifact,above] {$\mathcal{D}$} (graph.west);
    \draw[flow] (graph.east) -- node[artifact,above] {$G$} (check.west);
    \draw[flow] (check.east) -- node[artifact,above] {$M$} (output.west);
  \end{tikzpicture}
  \caption{\tech{} workflow. Arrows name inputs and intermediate artifacts.}
  \label{fig:architecture}
\end{figure*}
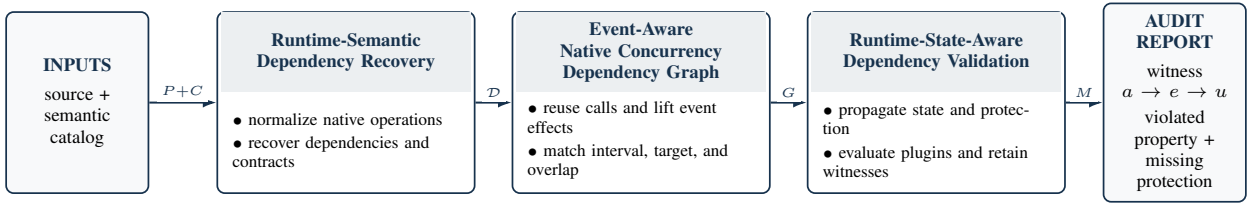

Figure~\ref{fig:architecture} shows three analysis components.
\emph{Runtime-Semantic Dependency Recovery} (D1) turns C
operations and API contracts into explicit dependencies and live regions.
The \emph{Event-Aware Native Concurrency Dependency Graph} (\ncdg{}, D2)
connects dependencies to interprocedural events through
interval, target identity, and overlap context. \emph{Runtime-State-Aware
Dependency Validation} (D3) propagates target-matched effects
and property-specific protection through live regions. Six mechanism plugins
select family-specific obligations while sharing this reasoning. In the figure,
$P$ and $C$ are the native program and catalog; $\mathcal{D}$, $G$, and $M$ are
recovered dependencies, the \ncdg{}, and candidates.
The data flow can be summarized as
\[
(P,C) \xrightarrow{\mathcal{R}} (F,\mathcal{D},\Pi,CG,S)
\xrightarrow{\mathcal{G}} G
\xrightarrow{\mathcal{V}_{\mathcal{P}}} M,
\]
Here $P$ is the native program, $C$ is the semantic catalog, and $F$ is the
normalized function representation. $\mathcal{D}$ is the dependency set.
$\Pi$, $CG$, and $S$ are the shared points-to result, call graph, and summary
fixed point. $G$ is the event-aware \ncdg{}, $\mathcal{P}$ is the plugin set,
and $M$ is the candidate set. For the running example, the first boundary
carries $D_{view}$. The second links it to release event $e$ through interval,
target, and overlap witnesses. The third reports that the required lease may be
false at use $u$.

\subsection{Problem Formulation}\label{sec:problem}

A native dependency is
\begin{equation}
D = \langle f,a,U,o,s,r,h,w,I,K,P\rangle,
\end{equation}
where $f$ is a function, $a$ is an acquisition, and $U$ contains later uses.
$o$ and $s$ are the owner and subject, and $r$ is their relation. $h$ and $w$
record the dependency family and acquisition ownership. $I$ is the live region
from acquisition to the relevant uses, $K$ is the stability contract, and $P$
records protection. For example, a list-item borrow relates
the list owner to an unowned item subject.  Its lifetime obligation requires
the item to remain live; a distinct relation obligation applies when the later
use relies on the item still occupying that owner role.

A concurrency event is
\begin{equation}
E = \langle f_e,p,t,c,X\rangle,
\end{equation}
where $f_e$ is the event's enclosing function, $p$ is its program point, $t$ is
the event target, $c$ is its execution context, and $X$ contains its possible
invalidation effects. Effects include
removing a relation, destroying a subject, resizing storage, rebuilding a
representation, releasing a lease, closing a resource, and re-entering code
that may mutate state.

For a risk class $R$ with required contract $K_R$, \tech{} reports
$\langle D,E,R\rangle$ only when
\begin{equation}\label{eq:candidate}
\begin{aligned}
  &\mathsf{Between}(E,I) \land \mathsf{MayTarget}(E,D) \\
  &\land\ \mathsf{MayOverlap}(E,D)\\
  &\land\ \exists k\in K_R,\,u\in U:
     \neg\mathsf{Protected}(D,E,k)\\
  &\hspace{18mm}{}\land\mathsf{ReachInvalid}(E,u,k).
\end{aligned}
\end{equation}
Here $\mathsf{ReachInvalid}(E,u,k)$ holds when the modeled effect of $E$ makes
property $k$ possibly invalid and propagation carries that state to a
consumption at $u$, without a kill, applicable discharge, or stabilizing
reacquisition. Section~\ref{sec:runtime-state} defines the placement of local
events and external-entry alternatives. The same $k$ indexes
protection, invalidation, and use: an event effect is suppressed only when the
recognized guard preserves that property over the complete dependency
interval. The first three terms are independently derived may facts. Their
conjunction forms an \emph{abstract dependency-invalidation witness}. Source
audit and dynamic or causal qualification then check whether one concrete
execution realizes the target, context, and failure.

\tech{} emits only state-violation candidates satisfying
Equation~\ref{eq:candidate}, retaining dependency, exposure, target, interval,
state, and protection evidence for inspection. The controlled ablation in
Section~\ref{sec:rq1} selectively disables these checks to measure their
contribution. User-visible reports still follow Equation~\ref{eq:candidate}.

\noindent\textbf{Scope of the reporting criterion.}
Equation~\ref{eq:candidate} defines the reporting rule. Coverage depends on
the compilation database, semantic catalog, call graph, points-to result, event
model, and protection abstraction. Ordinary transfers and events compose across
control-flow graph (CFG) points, but acquisition-time external-entry summaries
are single-event alternatives, without closure over multiple events at that
point. Missing catalog facts, incompatible may facts, unmodeled event sequences,
conditional builds, inline assembly, and external code can cause missed or
infeasible reports. A report-free result means that the supplied model produced
no candidate.

\subsection{Runtime-Semantic Dependency Recovery}
\label{sec:semantic-lifting}

Given native source $P$ and a semantic catalog $C$, this stage recovers
each acquisition's owner, subject, relation, stability contract, later
uses, and live region. It first normalizes runtime operations and computes
shared interprocedural facts, then constructs dependencies with the
source evidence needed by subsequent event connection and validation.

\subsubsection{Runtime-Semantic Normalization}
\label{sec:semantic-normalization}

Normalization translates C syntax into runtime-semantic operations while
preserving control flow, value identity, and source locations. Its goal
is to give macros, helpers, and API calls a consistent interpretation
that dependency recovery and event connection can share.

\noindent\textbf{Procedure.}
Algorithm~\ref{alg:dependency-recovery} normalizes each function and applies
catalog semantics
(lines~\ref{ln:recovery-normalize}--\ref{ln:recovery-collect}),
then computes the shared interprocedural facts
(lines~\ref{ln:recovery-pta}--\ref{ln:recovery-summaries}).

\SetAlCapFnt{\scriptsize}
\SetAlCapNameFnt{\scriptsize}
\begin{algorithm}[!tb]
\scriptsize
\setlength{\algomargin}{0.35em}
\DontPrintSemicolon
\LinesNumbered
\caption{Recover Native Dependencies}
\label{alg:dependency-recovery}

\SetKwFunction{Lift}{normalizeSource}
\SetKwFunction{Catalog}{applyCatalogSemantics}
\SetKwFunction{Uses}{reachableUses}
\SetKwFunction{Region}{liveRegion}
\SetKwFunction{Contract}{deriveContract}
\SetKwFunction{PTA}{pointsToAnalysis}
\SetKwFunction{Calls}{resolveCallGraph}
\SetKwFunction{Summaries}{propagateSummaries}
\SetKw{Return}{return}

\KwIn{$P$: native program and CFGs; $C$: semantic catalog}
\KwOut{$F$: normalized functions; $\mathcal{D}$: native dependencies;
$\Pi,CG,S$: shared interprocedural facts}

$F\leftarrow\emptyset$; $\mathcal{D}\leftarrow\emptyset$\;
\ForEach{function $f\in P$}{
    $F_f\leftarrow\Lift(f)$
        \nllabel{ln:recovery-normalize}\;
    $F_f\leftarrow\Catalog(F_f,C)$
        \nllabel{ln:recovery-catalog}\;
    $F\leftarrow F\cup\{F_f\}$
        \nllabel{ln:recovery-collect}\;
}
$\Pi\leftarrow\PTA(F)$; $CG\leftarrow\Calls(F,\Pi)$
    \nllabel{ln:recovery-pta}\;
$S\leftarrow\Summaries(F,CG,\Pi)$
    \nllabel{ln:recovery-summaries}\;

\ForEach{function $F_f\in F$}{
    \ForEach{semantic acquisition $a\in F_f$}{
        $U\leftarrow\Uses(a.subject,F_f,\Pi,CG,S)$
            \nllabel{ln:recovery-uses}\;
        $U\leftarrow\{u\in U\mid
          \mathrm{pathWithoutRedefinition}(a,u)\}$
            \nllabel{ln:recovery-filter}\;
        \If{$U=\emptyset$}{continue\;}

        $I\leftarrow\Region(a,U,F_f)$
            \nllabel{ln:recovery-region}\;
        $(o,s,r,h,w,K,P_a)\leftarrow\Contract(a,C)$
            \nllabel{ln:recovery-contract}\;
        $\mathcal{D}\leftarrow\mathcal{D}\cup
          \{\langle F_f.id,a,U,o,s,r,h,w,I,K,P_a\rangle\}$
            \nllabel{ln:recovery-emit}\;
    }
}
\Return{$(F,\mathcal{D},\Pi,CG,S)$}\;
\end{algorithm}

The frontend maps relevant Clang abstract syntax tree (AST) and CFG
nodes to object and interior-storage acquisition, traversal, leases,
uses, promotion, reference drops, ownership transfer, mutation, and
resource release. It also captures protection entry and exit, Python
re-entry, unknown calls, GIL changes, and thread creation and join.
Values retain runtime-oriented kinds such as object, storage, buffer,
iterator, callback, version, and guard.
The versioned catalog supplies acquisitions, ownership, owner--subject
relations, required stability, callbacks, effects, argument roles, and
protection. Rules describe reusable semantics, such as a borrowed
list-element return, independently of source-site labels. Explicit API
semantics override generic type and source-pattern rules; compatible
effects are unioned, while contradictory ownership declarations fail
validation.

We construct the catalog in four passes. First, CPython declarations,
C-API documentation, and \texttt{refcounts.dat} supply ownership facts
for new, borrowed, and stolen references~\cite{py-reference-ownership}.
Second, we inspect macros and source-local helpers exercised by the
compilation database for interior addresses, output-pointer aliases,
container relations, and mutation effects. Third, Argument Clinic
metadata and lock, critical-section, buffer-export, GIL, and callback
primitives supply protection and exposure facts. Finally, release diffs
identify changed boundary operations~\cite{cpython-source}. Validation
checks the schema, ownership consistency, duplicates, and named-rule
occurrences. Coverage audits separately count exercised named rules
and acquisitions in the qualified root corpus. Unmodeled helpers and
representation conventions remain outside scope until normalization
or catalog rules describe them.

After catalog application, \tech{} computes the points-to result $\Pi$,
resolved call graph $CG$, and summary fixed point $S$. Recovery uses
them to follow aliases and calls; NCDG construction reuses them to
connect dependencies to events. Figure~\ref{fig:architecture} therefore
separates semantic responsibilities while the implementation shares
their prerequisite analyses.

\noindent\textbf{Rationale.}
C types and expanded macros expose structure but not ownership:
neither \texttt{PyObject*} nor expanded \texttt{PyList\_GET\_ITEM}
identifies a result as borrowed from a particular
list~\cite{py-free-threading-extensions}. Normalization preserves the
program's control and data identities, while the catalog supplies the
missing runtime contracts. This separation gives all plugins consistent
semantics and makes release changes explicit rather than duplicating
API-name rules across detectors.

\subsubsection{Dependency and Live-Region Construction}
\label{sec:dependency-construction}

Dependency construction connects each acquisition to the uses that
consume its subject or derived aliases. It produces explicit contracts
and acquisition-to-use regions, so subsequent stages can determine
which events may invalidate a property before it is consumed.

\noindent\textbf{Procedure.}
Algorithm~\ref{alg:dependency-recovery} finds direct and summary-derived
uses through $\Pi$, $CG$, and $S$, retaining paths without subject
redefinition
(lines~\ref{ln:recovery-uses}--\ref{ln:recovery-filter}).
Acquisitions without later uses are omitted. The live region $I$
records blocks and instruction ranges on the remaining paths
(line~\ref{ln:recovery-region}). Subject assignment, fresh acquisition,
stable copying, owned-reference promotion, or modeled release can end
the corresponding path after that operation's own uses. Actual--formal
bindings and summaries carry uses and kills across calls; loop
reacquisition starts a new dependency.

Acquisition semantics supply the owner $o$, subject $s$, relation $r$,
family $h$, ownership $w$, configured contract $K$, and protection
$P_a$ (line~\ref{ln:recovery-contract}). These fields, the acquisition,
uses, and live region form the dependency
(line~\ref{ln:recovery-emit}). Mechanism plugins select their required
properties, while normalized use roles specify consumption before or
after attached interference.
For borrowed objects, the evaluated plugin conservatively requires
both \textsc{Live}, meaning subject lifetime, and \textsc{Rel},
meaning that the acquisition's owner--subject relation still holds.
A separately pinned object can be removed safely before a lifetime-only
use under a snapshot contract, although this default may emit a
relation candidate for source audit. A use that assumes the saved
subject still denotes the owner's selected entry requires both
properties.

\noindent\textbf{Rationale.}
Alias and summary information connects acquisitions to uses beyond
their original variables and functions. CFG-derived live regions
distinguish branch-specific uses, killed dependencies, and fresh loop
acquisitions that lexical intervals would merge. Keeping lifetime,
relation, and storage obligations separate prevents owner liveness
from being mistaken for protection of every dependent property.

\noindent\textbf{Running example---recovered dependency.}
For Listing~\ref{fig:motivation}, normalization and catalog application
identify \texttt{VIEW\_ADDR(self)} at $a$ as obtaining a pointer to
\texttt{self.view}, the memoryview's private \texttt{Py\_buffer}
descriptor~\cite{py-memoryview-api}. Points-to and field-path recovery
retain the chain from \texttt{self} through its managed buffer to the
exporter. The later \texttt{PyBuffer\_ToContiguous} call at $u$
consumes the descriptor and reads \texttt{self.view.buf}, requiring
an open view, an active lease, and a stable backing address.
Recovery instantiates Section~\ref{sec:problem}'s dependency schema
as $D_{view}$:
\[
\begin{aligned}
(f,a,U) &= (\texttt{memoryview\_tobytes},a,\{u\}),\\
(o,s,r) &= (\texttt{self},\texttt{self.view},\mathsf{BackingView}),\\
(h,w,I) &= (\mathsf{ViewLease},\mathsf{borrowed},I_{a\leadsto u}),\\
(K,P) &= (K_{view},\emptyset),
\end{aligned}
\]
where
$K_{view}=\{\mathsf{ViewOpen},\mathsf{LeaseActive},
\mathsf{AddressStable}\}$.
The active call keeps \texttt{self} alive but does not preserve these
view properties. \texttt{CHECK\_RELEASED} establishes
$\mathsf{ViewOpen}$ only at the check, not throughout the copy.
The protection set is empty because \texttt{tobytes} neither
increments \texttt{self->exports} nor holds another lease across
the acquisition-to-use interval. The next stage connects $D_{view}$
to events that may affect the same view.
%before the copy completes.

\subsection{Event-Aware Native Concurrency Dependency Graph}
\label{sec:ncdg}

Given normalized functions, recovered dependencies, and shared
interprocedural facts, this stage builds an event-aware \ncdg{}.
It identifies events that may execute within a dependency's live
region, affect a compatible owner or subject, and overlap through
parallel execution or re-entry. All mechanism plugins share the
resulting graph.

\begin{figure*}[t]
\centering
\begin{tikzpicture}[
  entry/.style={draw,rounded corners,align=center,text width=28mm,
                minimum height=10mm,fill=black!4,font=\scriptsize},
  op/.style={draw,rounded corners,align=center,text width=31mm,
             minimum height=11mm,fill=blue!4,font=\scriptsize},
  event/.style={draw,rounded corners,align=center,text width=40mm,
                minimum height=13mm,fill=red!5,font=\scriptsize},
  target/.style={draw,rounded corners,align=center,text width=42mm,
                 minimum height=13mm,fill=green!5,font=\scriptsize},
  cfg/.style={-{Latex[length=1.7mm]},thick},
  rel/.style={-{Latex[length=1.7mm]},thick,dashed},
  node distance=8mm and 8mm]

  \node[entry] (entryA) {public entry A\\\texttt{tobytes}};
  \node[op,right=of entryA] (acquire) {$a$: \textsc{AcquireView}\\
    \texttt{src = VIEW\_ADDR(self)}};
  \node[op,right=of acquire] (check) {point fact\\
    \texttt{CHECK\_RELEASED}\\\texttt{(self)}};
  \node[op,right=of check,text width=35mm] (use) {$u$: \textsc{UseBacking}\\
    \texttt{PyBuffer\_ToContiguous}};
  \draw[cfg] (entryA) -- (acquire);
  \draw[cfg] (acquire) -- (check);
  \draw[cfg] (check) -- (use);
  \draw[blue!65!black,thick]
    ($(acquire.south west)+(0,-3mm)$) --
    node[below,font=\scriptsize] {$D_{view}$ live region $I_{a\leadsto u}$}
    ($(use.south east)+(0,-3mm)$);

  \node[entry,below=14mm of entryA] (entryB) {public entry B\\
    \texttt{release}};
  \node[event,below=14mm of acquire] (event) {$e$: \textsc{ReleaseView}\\
    effect: \textsc{LeaseActive}$\leftarrow\mathsf{false}$\\
    backing export may be released};
  \node[target,below=14mm of use] (target) {canonical target\\
    \texttt{self.view}$\rightarrow$managed buffer\\$\rightarrow$exporter storage};
  \draw[cfg] (entryB) -- (event);

  \draw[rel,purple!75!black,<->]
    (entryA.south) --
    node[left,align=right,font=\scriptsize]{\textsc{MayOverlap}\\free-threaded entries}
    (entryB.north);
  \draw[rel,red!70!black]
    (event.north) --
    node[right=2mm,align=left,font=\scriptsize,fill=white,inner sep=1pt]
      {\textsc{MayExecuteBetween}\\may occur in $I_{a\leadsto u}$}
    ($(acquire.south)+(0,-3mm)$);
  \draw[rel,green!45!black] (event) --
    node[above,font=\scriptsize,fill=white,inner sep=1pt]
      {\textsc{MayTarget}} (target);
  \draw[rel,green!45!black] (use.south) --
    node[right=2mm,align=left,font=\scriptsize]{owner/subject\\path} (target.north);
\end{tikzpicture}
\caption{The running example's NCDG fragment. Solid arrows show native
control flow or summarized call flow; dashed arrows show NCDG relations.
The graph connects the release effect to the backing-buffer use through
the dependency's live region and canonical runtime target.}
\label{fig:running-ncdg}
\end{figure*}
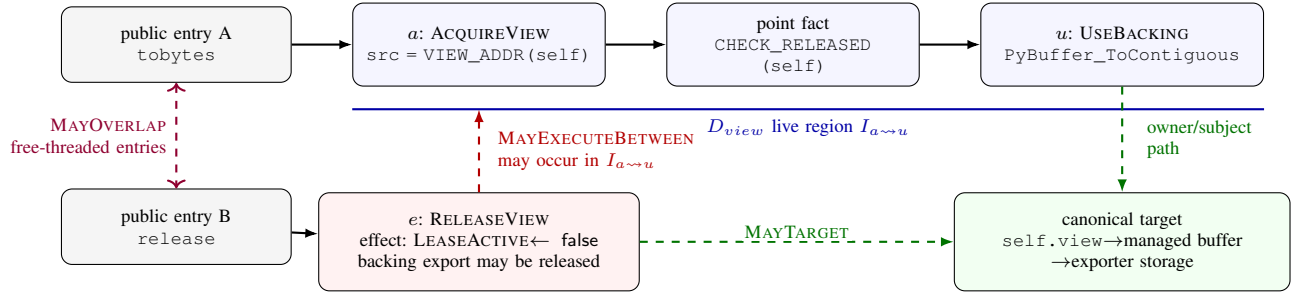

\subsubsection{Event and Context Construction}
\label{sec:ncdg-events}

The graph combines normalized program points, call and return edges,
dependencies, events, canonical targets, and summaries. Events retain
their effects and execution contexts so that dependency matching can
connect native uses to local and interprocedural invalidations.

\noindent\textbf{Procedure.}
Algorithm~\ref{alg:ncdg-construction} reuses the points-to result $\Pi$,
resolved call graph $CG$, and summary fixed point $S$ from
Algorithm~\ref{alg:dependency-recovery}. It constructs the interprocedural
CFG, lifts events, and indexes them by canonical target
(lines~\ref{ln:ncdg-cfg}--\ref{ln:ncdg-index}), then initializes the graph
(line~\ref{ln:ncdg-init}).

\SetAlCapFnt{\scriptsize}
\SetAlCapNameFnt{\scriptsize}
\begin{algorithm}[!tb]
\scriptsize
\setlength{\algomargin}{0.35em}
\DontPrintSemicolon
\LinesNumbered
\caption{Construct the Event-Aware \ncdg{}}
\label{alg:ncdg-construction}

\SetKwFunction{InterCFG}{buildInterCFG}
\SetKwFunction{LiftEvents}{liftEvents}
\SetKwFunction{Index}{indexByCanonicalTarget}
\SetKw{Return}{return}

\KwIn{$F,\mathcal{D},\Pi,CG,S$: output and shared facts from
Algorithm~\ref{alg:dependency-recovery}}
\KwOut{$G$: event-aware \ncdg{}}

$G_{cfg}\leftarrow\InterCFG(F,CG)$
    \nllabel{ln:ncdg-cfg}\;
$X\leftarrow\LiftEvents(F,S)$
    \nllabel{ln:ncdg-events}\;
$J\leftarrow\Index(X,S)$
    \nllabel{ln:ncdg-index}\;
$G\leftarrow\langle G_{cfg},\mathcal{D},X,S\rangle$
    \nllabel{ln:ncdg-init}\;

\ForEach{dependency $D\in\mathcal{D}$}{
    \ForEach{event $E\in J[\mathrm{targets}(D)]$}{
        \If{$\mathrm{mayExecuteBetween}(E,D,G_{cfg})$}{
            $G\leftarrow G\cup
              \{(D,E,\textsc{MayExecuteBetween})\}$
                \nllabel{ln:ncdg-between}\;
        }
        \If{$\mathrm{mayTarget}(E,D,\Pi,S)$}{
            $G\leftarrow G\cup
              \{(D,E,\textsc{MayTarget})\}$
                \nllabel{ln:ncdg-target}\;
        }
        \If{$\mathrm{mayOverlap}(E,D,G_{cfg},S)$}{
            $G\leftarrow G\cup
              \{(D,E,\textsc{MayOverlap})\}$
                \nllabel{ln:ncdg-overlap}\;
        }
    }
}
\Return{$G$}\;
\end{algorithm}

The interprocedural CFG preserves call, return, and intraprocedural
reachability. Summaries record effects on formal arguments and
canonical fields, escaping arguments, and exposure contexts; their
propagated bindings associate formal targets with actual objects or
fields. Unknown callees conservatively expose pointer-like arguments
and carry cataloged effects.
Events include mutation, reference drops, ownership transfer,
resource release, callback and Python re-entry, finalization, blocking
and foreign calls, lock operations, thread creation and join, and GIL
transitions. Contexts distinguish the native continuation, Python
threads, callbacks, foreign threads, finalizers, and garbage collection.
The target index groups events by canonical objects or field paths 
for subsequent dependency matching.

\noindent\textbf{Rationale.}
A dependency and its invalidation may occur in different functions.
Summary bindings connect callee effects to caller objects, while the
interprocedural CFG preserves the paths needed to relate those effects
to later uses. Canonical-target indexing limits matching to compatible
event buckets rather than an all-to-all dependency--event product.

\subsubsection{Dependency--Event Connection}
\label{sec:ncdg-relations}

For each indexed dependency--event pair, the builder independently
records \textsc{MayExecuteBetween}, \textsc{MayTarget}, and
\textsc{MayOverlap}. The next stage considers the event eligible for
state validation only when all three relations hold.

\noindent\textbf{Procedure.}
\textsc{MayExecuteBetween}
(line~\ref{ln:ncdg-between}) follows acquisition-to-use reachability.
An intraprocedural event requires an acquisition--event--use CFG path
without a dependency kill. An external-entry summary is eligible for
an acquisition-time alternative only after the target may have
escaped to that entry.
\textsc{MayTarget}
(line~\ref{ln:ncdg-target}) holds when values match directly, their
points-to sets intersect, actual--formal bindings match, or canonical
field paths share a root and compatible suffix. This connects an event
to the dependency's runtime target rather than merely to a nearby
source operation.
\textsc{MayOverlap}
(line~\ref{ln:ncdg-overlap}) uses public-entry exposure, thread
creation and join, callback and finalizer contexts, blocking calls,
and GIL state. Free-threaded public entries may overlap unless
established ordering or a join excludes them. Cataloged callbacks,
finalizers, foreign threads, signals, and GIL-release contexts can
also establish overlap within the live region. Conventional-GIL
mode requires these explicit exposure contexts rather than arbitrary
public-entry overlap.

Protection constrains eligible event effects. A lock or critical
section suppresses an effect only when its guard is established as
compatible, every path from acquisition through the affected uses
retains protection, and the invalidating operation follows the same
synchronization protocol. Guard--target may-aliasing alone is
insufficient. Strong references, buffer exports, private phases,
version checks, and lifecycle rules suppress only effects on the
properties they preserve; an unprotected contract-relevant effect
remains eligible for transfer.
CPython may suspend an outer critical section when entering another
or detaching thread state~\cite{py-critical-sections}. The prototype
recognizes lexical entry and exit regions and explicit callback,
blocking, and GIL-release events, but not the complete suspension
and resumption stack. Unmodeled suspension can therefore overstate
protection and cause false negatives.

\noindent\textbf{Rationale.}
Invalidation can cross memory locations: dictionary removal may
reclaim an object later incremented elsewhere, and view release may
revoke storage later copied. The \ncdg{} connects these semantic
effects to dependencies without requiring a shared accessed address.
Separate interval, target, and overlap relations distinguish when
an event may occur, what it may affect, and which execution context
permits interference. All six plugins reuse this call, alias, and
concurrency evidence. These relations remain conservative may facts;
their conjunction establishes eligibility, not a concrete violation.

\noindent\textbf{Running example---NCDG fragment.}
Figure~\ref{fig:running-ncdg} shows the graph for $D_{view}$.
The upper path is the native CFG slice of
\texttt{memoryview.tobytes}; the lower summarizes the peer public
\texttt{release} entry. Event $e$ carries the effect
$\mathsf{LeaseActive}\leftarrow\mathsf{false}$ and may drop the
backing export associated with \texttt{self.view}.
The graph records three connections: $e$ may occur in
$I_{a\leadsto u}$, its canonical target matches the view's backing
relation, and the two public entries may overlap in free-threaded
execution. The PoC's later resize exercises a backing-storage change
enabled by lease loss; release itself revokes the lease required
at $u$.
If any of the three relations is absent, $e$ is not eligible for
validation of $D_{view}$. A release ordered after $u$ fails
\textsc{MayExecuteBetween}; one with a disjoint target fails
\textsc{MayTarget}; and ordinary conventional-GIL entries fail
\textsc{MayOverlap} without explicit re-entry or GIL release.
The resulting graph supplies the dependency and its eligible release
event to the state-validation stage.

\subsection{Runtime-State-Aware Dependency Validation}
\label{sec:runtime-state}

Given the \ncdg{} and mechanism contracts, this stage checks whether
an eligible event invalidates a property required by a later use.
It produces an abstract witness linking the dependency, the
unprotected event effect, and the use that observes invalid state.
A shared core handles state propagation and protection; six plugins
select contracts and classify the resulting violations.

\subsubsection{Property-Specific State Propagation}
\label{sec:state-propagation}

The core analyzes each dependency over its live CFG region. It tracks
the properties required by applicable plugins, applies local operations
and event effects, and records violations that reach a use.

\SetAlCapFnt{\footnotesize}
\SetAlCapNameFnt{\footnotesize}
\begin{algorithm}[t]
\footnotesize
\setlength{\algomargin}{0.35em}
\DontPrintSemicolon
\LinesNumbered
\caption{Detect Dependency Invalidation}
\label{alg:state-checking}

\SetKwFunction{Applicable}{applicablePlugins}
\SetKwFunction{Init}{postAcquisitionState}
\SetKwFunction{External}{externalAlternatives}
\SetKwFunction{Requires}{requiredOnInput}
\SetKwFunction{PostConsumes}{postUse}
\SetKwFunction{Post}{localPostTransfer}
\SetKwFunction{Events}{eligibleEvents}
\SetKwFunction{Effects}{unprotectedEffects}
\SetKwFunction{Transfer}{transfer}
\SetKwFunction{JoinState}{joinState}
\SetKwFunction{Violations}{violations}
\SetKwFunction{Record}{recordTransition}
\SetKwFunction{Evaluate}{evaluate}
\SetKwFunction{Candidate}{buildCandidate}
\SetKw{Return}{return}

\KwIn{$G$: event-aware \ncdg{}; $\mathcal{P}$: mechanism plugins}
\KwOut{$M$: state-violation candidates}

$M\leftarrow\emptyset$\;
\ForEach{dependency $D\in G.\mathcal{D}$}{
    $P_D\leftarrow\Applicable(D,\mathcal{P})$
        \nllabel{ln:state-plugins}\;
    \If{$P_D=\emptyset$}{continue\;}
    $K_D\leftarrow
      \bigcup_{p\in P_D}p.\mathrm{requiredStability}(D)$
        \nllabel{ln:state-contract}\;
    $A[\cdot]\leftarrow\bot$;
    $\sigma_a\leftarrow\Init(D)$
        \nllabel{ln:state-init}\;
    $(\sigma_a,T)\leftarrow\External(D,\sigma_a,K_D)$
        \nllabel{ln:state-external}\;
    $A[D.a]\leftarrow\sigma_a$\;
    $W\leftarrow\{D.a\}$; $V_D\leftarrow\emptyset$\;

    \While{$W\neq\emptyset$}{
        $q\leftarrow\mathrm{pop}(W)$;
        $\sigma\leftarrow A[q]$\;
        $K_{in}\leftarrow\Requires(D,q)$
            \nllabel{ln:state-input}\;
        \If{$K_{in}\neq\emptyset$}{
            $V_{in}\leftarrow\Violations(\sigma,K_{in})$\;
            \ForEach{$k\in V_{in}$}{
                $V_D\leftarrow V_D\cup\{(q,k)\}$
                    \nllabel{ln:state-input-record}\;
            }
        }
        $\sigma_0\leftarrow\Post(D,q,\sigma)$
            \nllabel{ln:state-local}\;
        $S\leftarrow\{\sigma_0\}$\;
        \ForEach{$E\in\Events(D,q)$}{
            $X\leftarrow\Effects(D,E,\sigma_0,K_D)$
                \nllabel{ln:state-protection}\;
            \If{$X=\emptyset$}{continue\;}
            $\sigma_E\leftarrow\Transfer(\sigma_0,E,X)$;
            $\Delta_E\leftarrow\Violations(\sigma_E,K_D)
                       \setminus\Violations(\sigma_0,K_D)$
                \nllabel{ln:state-transfer}\;
            $T\leftarrow\Record(T,E,\Delta_E,\sigma_0,\sigma_E)$
                \nllabel{ln:state-witness}\;
            $S\leftarrow S\cup\{\sigma_E\}$\;
        }
        \ForEach{$\sigma'\in S$}{
            \If{$\PostConsumes(D,q)$}{
                $V\leftarrow\Violations(\sigma',K_D)$
                    \nllabel{ln:state-post-use}\;
                \ForEach{$k\in V$}{
                    $V_D\leftarrow V_D\cup\{(q,k)\}$\;
                }
            }
            \If{$q$ keeps the current instance of $D$ live}{
                \ForEach{successor $q'$ in $D.I$}{
                    $\hat{\sigma}\leftarrow
                        \JoinState(A[q'],\sigma')$
                            \nllabel{ln:state-join}\;
                    \If{$\hat{\sigma}\neq A[q']$}{
                        $A[q']\leftarrow\hat{\sigma}$;
                        $W\leftarrow W\cup\{q'\}$\;
                    }
                }
            }
        }
    }

    \ForEach{plugin $p\in P_D$}{
        $K_p\leftarrow p.\mathrm{requiredStability}(D)$\;
        $V_p\leftarrow\{k\mid(q,k)\in V_D\land k\in K_p\}$\;
        $T_p\leftarrow\mathrm{restrict}(T,V_p)$
            \nllabel{ln:state-restrict}\;
        $d\leftarrow\Evaluate(p,D,T_p)$
            \nllabel{ln:state-evaluate}\;
        \If{$d.\mathrm{emit}$}{
            $M\leftarrow M\cup\{\Candidate(D,p,T_p,d)\}$
                \nllabel{ln:state-report}\;
        }
    }
}
\Return{$M$}\;
\end{algorithm}

\noindent\textbf{Procedure.}
Algorithm~\ref{alg:state-checking} selects applicable plugins and
unions their required properties
(lines~\ref{ln:state-plugins}--\ref{ln:state-contract}).
It then runs one forward worklist for that dependency.
Here, $A$ stores analysis information at program points, $W$ is the
worklist, $T$ stores event transitions, and $V_D$ stores properties
found invalid at uses. CFG entries contain only the abstract state
$\sigma$.
The product state records relation validity, owner and subject
lifetime, reference ownership and count, storage address and content,
representation validity, lease and resource state, version, privacy,
lock depth, and GIL state. Validity uses
$\{\mathsf{Unknown},\mathsf{Valid},\mathsf{MayInvalid},\mathsf{Invalid}\}$;
lifetime uses
$\{\mathsf{Unknown},\mathsf{Live},\mathsf{MayDead},\mathsf{Dead}\}$;
and reference count uses
$\{\mathsf{Unknown},0,1,\mathsf{Many},\mathsf{Nonzero},\mathsf{MayZero}\}$.
Ownership is unknown, borrowed, owned, pinned, unowned, released, or
dead. Privacy, owner-held, and GIL-release facts are Boolean.
Lock depth records the nonnegative depth guaranteed on every incoming
path. Entry increments it, exit decrements it, and joins take the
minimum. The counter has no fixed cap; a loop backedge cannot increase
the stored guaranteed depth. Joins preserve agreement and widen
conflicting validity, lifetime, reference-count, and ownership values
to may or unknown states. The worklist repeats until joined states
stop changing.

Table~\ref{tab:state-transfers} summarizes the transfers. Each operation
changes only the properties it affects.

\begin{table}[!htb]
\centering
\caption{Principal abstract-state transfers. Other state properties
remain unchanged.}
\label{tab:state-transfers}
\scriptsize
\setlength{\tabcolsep}{3pt}
\begin{tabular}{@{}p{.27\columnwidth}p{.29\columnwidth}p{.39\columnwidth}@{}}
\toprule
Normalized operation & State update & Ordering or kill rule \\
\midrule
Dependency acquisition &
Required properties valid in the post-state &
Starts a new instance; the acquisition is not applied twice \\
Ownership promotion &
Lifetime/refcount $\leftarrow$ live/nonzero &
Checks the incoming borrow first; discharges only its borrowed-object
obligation \\
Redefinition or reacquisition &
Old instance killed &
Later uses belong to the new acquisition \\
Lock/GIL enter or leave &
Lock depth or GIL state updated &
Local transfer precedes attached events \\
Interval guard &
No repair of invalid state &
Suppresses covered effects over the full dependency interval \\
Relation removal/drop &
Relation may invalidate; unowned subject may die &
Ownership protects lifetime, not the relation \\
Resize/rebuild &
Address, representation, and version may invalidate &
An active lease may protect only the address \\
Release/close/re-entry &
Lease, resource, or callback state may invalidate &
Applies effect-specific protection first \\
\bottomrule
\end{tabular}
\end{table}

Initialization installs the post-acquisition state at $D.a$
(line~\ref{ln:state-init}). Processing that acquisition is a no-op;
loop reacquisition starts a new instance.
\textsf{externalAlternatives} joins the no-event state with one
alternative for each eligible external-entry event
(line~\ref{ln:state-external}). Each alternative represents the event
occurring after acquisition and before a later use. The engine
propagates its invalidation through the live region, subject to
property-specific discharges and interval protection. It forms no
two-event external sequence at this point. Local events remain at
their actual CFG operations and compose across distinct points;
effects within one event also compose.
At other program points, the state precedes the normalized operation.
\textsf{requiredOnInput} identifies properties consumed before
attached interference
(lines~\ref{ln:state-input}--\ref{ln:state-input-record}).
The core checks them before \textsf{localPostTransfer} changes
ownership, updates lock or GIL state, or kills the instance
(line~\ref{ln:state-local}).
Promotion therefore checks the borrowed input before setting lifetime
and reference count to live and nonzero. It discharges the
borrowed-object lifetime and relation obligations; separate storage,
lease, traversal, and callback dependencies remain.

Attached events follow the local transfer. A second check occurs
only when \textsf{postUse} marks consumption after interference
(line~\ref{ln:state-post-use}). A completed use has only an input
check. A callback-capable operation has a post-use check when its
native continuation consumes the retained dependency after callback
return; it may have both checks. An event after final consumption
does not produce a violation at that completed use.
An event is eligible only when all three \ncdg{} relations hold.
\textsf{unprotectedEffects} removes an effect only when recognized
protection preserves every property it could violate
(line~\ref{ln:state-protection}). Ownership and pinning protect
lifetime; immutability protects structure and storage. Locks and
critical sections require full-interval coverage of the relevant
target. The model also recognizes stable copies, active exports,
private construction, lifecycle exclusion, version guards, and
no-reentry regions. Version and no-reentry checks act as interval
guards rather than point transfers that repair invalid state.

Remaining effects update their state dimensions
(line~\ref{ln:state-transfer}). The core records an event when it
first makes a required property possibly invalid
(line~\ref{ln:state-witness}). A violation is collected only when
that state reaches an input or post-interference use.
Redefinition ends the current dependency and clears its state, so
fresh acquisitions start independently.
The CFG worklist stores only $\sigma$. A join widens abstract state
and does not select or merge witnesses. The separate ledger $T$
stores event transitions and their invalidated properties. After
propagation, the engine keeps ledger entries whose properties are
invalid at a use and occur in the plugin contract. This pairing is
conservative across may paths. Reports retain the contributing event
records for source review.

By default, the engine retains the first 16 external-entry witnesses
per violated property in deterministic traversal order. Further
external events still affect state but are not stored as witnesses.
Local witnesses are deduplicated by event ID without a fixed bound.
Section~\ref{sec:rq3} evaluates the configurable external bound.

\noindent\textbf{Rationale.}
Different guarantees preserve different properties. A strong reference
keeps an object alive but does not preserve an interior pointer across
resize. A buffer export stabilizes its address without freezing its
contents or protecting an unrelated traversal
~\cite{py-thread-safety-guarantees}.
Separate state dimensions retain these distinctions.
Checking inputs before promotion preserves the original obligation;
checking post-interference uses captures dependencies retained across
callbacks. Joins and instance kills keep the analysis aligned with
control flow and acquisition lifetime.

\subsubsection{Mechanism-Specific Reporting}
\label{sec:mechanism-reporting}

Plugins turn core-produced violations into mechanism-specific
candidates. They select dependencies and required properties, then
assign a review priority and explanation using the shared path,
target, overlap, protection, and state evidence.

\noindent\textbf{Procedure.}
A plugin is
\[
p_i=\langle id_i,Q_i,K_i,A_i,L_i\rangle,
\]
where $id_i$ names the mechanism, $Q_i(D,G)$ selects dependencies,
$K_i(D)$ returns required properties, $A_i(D,\Theta,G)$ decides
whether the transitions justify a report, and $L_i$ assigns priority
and rationale.
Let $\Theta_D$ contain the transitions reaching uses of $D$, with
their target, overlap, protection, state, and witness evidence.
After propagation, Algorithm~\ref{alg:state-checking} restricts these
transitions to each plugin's contract, evaluates them, and builds
accepted candidates
(lines~\ref{ln:state-restrict}--\ref{ln:state-report}):
\begin{tcolorbox}[tile,width=\linewidth,size=fbox,boxsep=1mm,boxrule=0.4pt,
                  before skip=5pt,after skip=5pt,
                  colback=black!2!white,colframe=black!55!white]
\scriptsize
\setlength{\abovedisplayskip}{2pt}
\setlength{\belowdisplayskip}{2pt}
\[
\begin{aligned}
\Theta_i(D)
  &=\{\theta\in\Theta_D\mid\theta.k\in K_i(D)\},\\
\textsc{Match}_i(D,\Theta_D)
  &\equiv Q_i(D,G)\land A_i(D,\Theta_i(D),G),\\
m
  &=\mathsf{Build}(D,id_i,\Theta_i(D),L_i(D,\Theta_i(D))).
\end{aligned}
\]
\end{tcolorbox}

The C++ methods \texttt{kind}, \texttt{applicable}, and
\texttt{requiredStability} implement $id_i$, $Q_i$, and $K_i$.
The \texttt{evaluate} method combines $A_i$ and $L_i$.
Table~\ref{tab:contracts} summarizes the six plugins.

\begin{table}[!htb]
\centering
\caption{Invalidation-mechanism contracts. All plugins use the core's
path, target, overlap, protection, and state evidence.}
\label{tab:contracts}
\scriptsize
\setlength{\tabcolsep}{3pt}
\begin{tabular}{@{}p{.23\columnwidth}p{.39\columnwidth}p{.33\columnwidth}@{}}
\toprule
Mechanism plugin & $Q_i$: dependency and effect & $K_i$: required property \\
\midrule
Borrowed-object lifetime &
Unowned subject; removal, replacement, or reference drop may destroy
or detach it &
Conservative default: subject remains live and its acquisition
relation holds until promotion or final use \\
Storage generation &
Raw or interior pointer; resize, reallocation, rebuild, close, or
free may revoke storage &
Address and representation remain valid through the last dereference \\
Container traversal &
Size, index, cursor, entry, or iterator; structural mutation may
invalidate its bounds, position, or relation &
Container structure and traversal state remain coherent through the
step \\
View/lease lifetime &
Exporter- or resource-backed view; release, close, detach, or unleased
resize may revoke backing &
Lease remains active, resource open, and address valid through use \\
Callback re-entry &
State retained across a callback-capable operation; callbacks,
finalizers, or foreign code may replace or clear it &
Pre-call relation or version holds after return, or a full-interval
guard covers the continuation \\
Reference-drop lifetime &
Object or field retained across a drop; drop-to-zero and destructor
re-entry may reclaim dependent state &
Subject remains live through the last access \\
\bottomrule
\end{tabular}
\end{table}

The view/lease plugin provides one instance of this interface.
$\mathsf{LeaseLike}(D)$ holds when $D$ is a lease relation or admits
a release or resource-close effect. $\mathsf{OpenIfRequired}(D)$
adds the openness property required by that dependency:
\begin{tcolorbox}[tile,width=\linewidth,size=fbox,boxsep=1mm,boxrule=0.4pt,
                  before skip=5pt,after skip=5pt,
                  colback=black!2!white,colframe=black!55!white]
\scriptsize
\setlength{\abovedisplayskip}{2pt}
\setlength{\belowdisplayskip}{2pt}
\[
\begin{aligned}
Q_V(D,G)
  &\equiv\mathsf{LeaseLike}(D)\land D.U\neq\emptyset,\\
K_V(D)
  &=\{\mathsf{LeaseActive},\mathsf{AddressStable}\}\\
  &\quad{}\cup\mathsf{OpenIfRequired}(D),\\
A_V(D,\Theta,G)
  &\equiv\exists\theta\in\Theta.\;
    \mathsf{Overlap}(\theta)\land\mathsf{MayContext}(\theta)\\
  &\quad{}\land\mathsf{TargetMatch}(\theta)
    \land\theta.\Delta\neq\emptyset
    \land\theta.V\neq\emptyset.
\end{aligned}
\]
\end{tcolorbox}
Here, $\theta.\Delta$ and $\theta.V$ contain the changed and violated
properties. The other predicates retain the core's overlap, context,
and target evidence. An accepted view/lease candidate receives the
highest review priority and names the release, close, detach, or
unleased resize that may revoke the backing before use.

Each candidate records acquisition, use, and event locations; owner
and subject hypotheses; relation and live-region precision; exposure
and protection evidence; changed and violated properties; and a
transition trace. These fields preserve
Equation~\ref{eq:candidate}'s reasoning for source audit and PoC
design. Mechanism labels describe the detected invalidation.
Confirmed roots receive separate primary Common Weakness Enumeration
(CWE) labels for security impact; several manifestations may share
one root.

\noindent\textbf{Rationale.}
Propagating the union of plugin contracts once gives every family
the same target and path evidence. Plugins specialize the selected
properties and report decisions without repeating propagation.
New families over existing properties reuse the core and \ncdg{};
new state dimensions or effects require extensions to the lattice
and transfer functions. Monotone transfers preserve the shared path
analysis, while witness retention controls the evidence stored
with reports.

\noindent\textbf{Running example---state transition and plugin result.}
For $D_{view}$, initialization records an open view, active lease,
and valid backing address. The \ncdg{} supplies release event $e$
with all three required relations. The active call keeps
\texttt{self} alive, but \texttt{CHECK\_RELEASED} adds no interval
lease and no temporary export covers $a\leadsto u$.
The core therefore applies $e$'s lease-revocation effect, records
$e$ as a witness, and propagates the possibly inactive lease to
the backing read at $u$. The view/lease plugin selects $K_{view}$,
accepts the lease violation, and emits a candidate with the witness
\[
a\xrightarrow{D_{view}}e\xrightarrow{\Delta K_{view}}u.
\]
An atomic check-and-pin that holds a temporary export through the
copy would protect the lease. The core would suppress the covered
release effect, distinguishing that interval guarantee from a point
check or a reference that keeps only \texttt{self} alive.

\section{Implementation}\label{sec:impl}

\tech{} is a 10.2K-line C++17 prototype, excluding tests and benchmarks.
CMake~3.20 builds two libraries and a command-line driver. The
LLVM/Clang~14 frontend exports normalized code; the analysis library
implements points-to analysis, call graphs, interprocedural summaries,
\ncdg{} construction, runtime-state checking, plugins, and report
serialization.

\noindent\textbf{Clang frontend.}
Clang Tooling loads \texttt{compile\_commands.json} and parses selected
translation units with their original flags~\cite{clang-compdb}.
The lifter visits declarations, expressions, macro expansions, and CFG
blocks~\cite{clang}. It assigns compact numeric identifiers to functions,
blocks, instructions, and values, preserves spelling and expansion
locations, and identifies address-taken functions for indirect-call
resolution.

\noindent\textbf{Analysis engines and memory management.}
Andersen-style inclusion analysis is the default~\cite{andersen1994};
a flow-sensitive implementation shares its program-point query
interface. Direct and resolved indirect calls form the call graph.
Delta worklists propagate effect, escape, and exposure summaries to
convergence. The implementation interns canonical symbols and target
sets and indexes external events by target.
The interprocedural CFG takes ownership of normalized functions,
allowing the driver to release the frontend copy before \ncdg{}
construction. Temporary points-to results and summaries are freed
after the graph copies the facts needed for state checking.

\noindent\textbf{Runtime-state engine and plugins.}
The product-lattice core implements
Algorithm~\ref{alg:state-checking} with a block worklist and stores
states only within the current dependency's live region. CFG joins
carry abstract state; a separate dependency-level ledger stores event
transitions. External-entry witness retention has a configurable
bound, while local provenance is deduplicated without a fixed bound.
Six plugins implement \texttt{kind}, \texttt{applicable},
\texttt{requiredStability}, and \texttt{evaluate}. They select
ledger records by property and apply the contracts in
Table~\ref{tab:contracts} using shared event and protection evidence.

\noindent\textbf{Semantic catalogs.}
The frontend loads versioned JSONL records for API summaries, macro
semantics, callbacks, effects, types, targets, and protection.
Free-threaded exposure is the default. Conventional-GIL mode requires
explicit re-entry, finalization, foreign-thread, blocking, or
GIL-release evidence.

\noindent\textbf{Driver and reports.}
The driver accepts a compilation database, optional translation-unit
subset, catalog, points-to mode, CPython commit, and diagnostic
switches. These switches control relation recovery, event connection,
protection policy, and the external-witness bound.
Versioned JSON reports contain normalized functions, dependencies,
events, graph relations, candidates, source evidence, and configuration
metadata. Dependency tags distinguish direct catalog instantiations,
catalog-derived wrapper relations, and source-inferred relations.
Diagnostics record witness retention and protection decisions; the
serialized graph preserves event IDs omitted from bounded reports.
A Linux profiler samples resident memory through
\texttt{/proc/self/statm} and records phase and end-to-end time.

\section{Evaluation}\label{sec:eval}

We test whether \tech{} distinguishes violating events from close semantic
controls, scales to a large production runtime, and exposes consequential
failures confirmable by executable tests:

\begin{description}[style=unboxed,leftmargin=0cm]
\denseitems
\item[\textbf{RQ1: Effectiveness.}] What precision/recall does \tech{} achieve
on ConcurrBench relative to controlled evidence configurations?
\item[\textbf{RQ2: Efficiency.}] What end-to-end and component time/memory does
whole-program CPython analysis require, compared with a flow-sensitive
points-to baseline?
\item[\textbf{RQ3: Real bugs.}] What findings does \tech{} confirm in production
CPython, what evidence establishes their consequences, and how do they vary
across releases and runtime configurations? Which external baselines accept and
complete their stated production inputs?
\end{description}

\noindent\textbf{Counting units.}
A \emph{candidate instance} is an emitted entry; Section~\ref{sec:impl}'s stable
fingerprint collapses duplicates. A \emph{confirmed manifestation} is a
fingerprint whose dependency/event survives source audit and dynamic or causal
qualification. A \emph{bug root} is a deduplicated native cause, possibly spanning
multiple manifestations; different stacks sharing a cause count once.
\emph{Confirmation yield} divides confirmed manifestations by unique emitted
fingerprints. Since remaining reports are not all classified as refuted or
unresolved, this production ratio is not precision.

\noindent\textbf{Targets and environment.}
We analyze every translation unit in the free-threaded (FT) compilation databases
for CPython 3.13.15 (\texttt{4061bc4}), 3.14.7 (\texttt{823f032}), and
3.15.0rc2 (\texttt{435c9e5})~\cite{cpython-source}. Version 3.13 introduced
experimental free threading~\cite{py313-free-threading}; 3.14 is our first
officially supported baseline~\cite{pep779}. Static runs use a 20-core Intel Core
Ultra 7 265K, 62\,GiB memory, Ubuntu 24.04.3, Clang~14, and Andersen. All static
compilation databases are free-threaded. Dynamic qualification distinguishes
\texttt{--disable-gil} builds with \texttt{PYTHON\_GIL=0}, the same build with
compatibility GIL enabled by \texttt{PYTHON\_GIL=1}, and separately compiled
conventional-GIL builds~\cite{py-free-threading-python}. Compatibility runs are
controls or labeled corroboration, not conventional-build evidence. Tests use
debug allocators/assertions, AddressSanitizer (ASan) where available~\cite{asan},
and semantic oracles where documented invariants exist.

The databases contain 284, 303, and 309 distinct C translation units, including
each interpreter's generated \texttt{Modules/config.c}. \tech{} lifts
14,191--15,518 functions, recovers 5,674--6,240 dependencies, and catalogs
92,369--104,282 events. Catalogs contain 338, 339, and 397 JSONL facts, including
294, 295, and 353 named API rules. Before analysis, schema validation rejects
unknown enums, inconsistent ownership/effects, duplicate API names, and malformed
roles. Catalogs follow Section~\ref{sec:semantic-lifting}'s four-pass procedure.
Validation makes inputs auditable; ConcurrBench supplies the explicit labeled
semantic-conformance oracle.

We audit rule utilization and qualified-root acquisition coverage separately.
Post-audit catalogs exercise 294/295, 293/296, and 351/354 named rules;
unused rules cover configuration-dependent source retained across releases.
Auditing 141 confirmed FT roots found two missing acquisitions: byte-array
storage returned through an internal helper's output parameter and an inherited
field with incompatible base/derived target paths. Adding a \texttt{format\_obj}
output relation and canonicalizing \texttt{PyOSErrorObject.args} to
\texttt{PyBaseExceptionObject.args} produces both candidates, reaching 141/141
qualified-root acquisition coverage. Scored yield retains frozen catalogs;
ConcurrBench remains the recall experiment. Released-catalog whole-program runs
emit 1,564, 1,473, and 1,514 instances versus 1,563, 1,472, and 1,512 frozen
instances. The one, one, and two additions map to these two omitted roots.

\subsection{RQ1: Effectiveness on ConcurrBench}\label{sec:rq1}

\noindent\textbf{Result.}
The full configuration classifies all 90 positive/negative pairs correctly.
All positives survive the cumulative configurations; exposure removes 23
controls, target/live-region checks remove 20, state/protection removes 44, and
local recovery rejects three cases with no live dependency.

\begin{table}[!t]
\centering
\caption{ConcurrBench semantic conformance and cumulative evidence ablations.  Each
plugin row contains 15 positive and 15 matched negative cases.  Ablation rows
cover the complete 180-case benchmark and cumulatively add evidence.  TP, FP,
FN, and TN denote true positives, false positives, false negatives, and true
negatives.}
\label{tab:concurrbench}
\small
\setlength{\tabcolsep}{2pt}
\begin{tabularx}{\columnwidth}{@{}Xrrrrrr@{}}
\toprule
Configuration/scope & TP & FP & FN & TN & Precision & Recall \\
\midrule
\multicolumn{7}{@{}p{\columnwidth}@{}}{\emph{Full \tech{} configuration, by mechanism plugin}} \\
Borrowed-object lifetime & 15 & 0 & 0 & 15 & 100\% & 100\% \\
Storage generation & 15 & 0 & 0 & 15 & 100\% & 100\% \\
Container traversal & 15 & 0 & 0 & 15 & 100\% & 100\% \\
View/lease lifetime & 15 & 0 & 0 & 15 & 100\% & 100\% \\
Callback re-entry & 15 & 0 & 0 & 15 & 100\% & 100\% \\
Reference-drop lifetime & 15 & 0 & 0 & 15 & 100\% & 100\% \\
\midrule
\multicolumn{7}{@{}p{\columnwidth}@{}}{\emph{Same-front-end evidence ablations, full benchmark}} \\
Dependency recovery only & 90 & 87 & 0 & 3  & 50.85\% & 100\% \\
$+$ exposure check & 90 & 64 & 0 & 26 & 58.44\% & 100\% \\
$+$ target and live-region checks & 90 & 44 & 0 & 46 & 67.16\% & 100\% \\
\textbf{$+$ state and protection checks} & \textbf{90} & \textbf{0} &
\textbf{0} & \textbf{90} & \textbf{100\%} & \textbf{100\%} \\
\bottomrule
\end{tabularx}
\end{table}

\noindent\textbf{Method.}
ConcurrBench comprises 180 C programs in 90 positive/negative pairs, 15 per plugin.
Each pair preserves control/data flow but changes one decisive property:
same/disjoint target, mutation before/after final use, borrowed/owned access,
partial/full protection, or stale use/reacquisition. Machine-readable manifests
fix expected dependency, event, target, overlap, protection, state, and label;
they were frozen before source-generator and scoring-script implementation.

\noindent\textbf{Controlled ablations.}
Table~\ref{tab:concurrbench} compares the full configuration by plugin with three
cumulative same-front-end ablations. Fixed inputs, frontend, dependency graph,
and oracle isolate exposure, target/interval, and state/protection evidence
admitted before reporting.

Matched controls test target identity, temporal overlap, stable ownership,
interval validation, and full-interval protection. They reject reports based
on nearby mutation alone. ConcurrBench isolates these reasoning dimensions; RQ3
measures production CPython behavior.

Five regression tests exercise Algorithm~\ref{alg:state-checking}'s ordering.
External alternatives survive acquisition processing and reach later uses.
Invalidation before promotion reports at promotion; later invalidation is
suppressed for the discharged borrow. Critical sections suppress covered
relation mutation while retaining unrelated lifetime drops. Loop reacquisition
starts a fresh instance. Paired consumption cases suppress post-final-use events
and report post-event continuation uses. ConcurrBench's denominator remains 180
cases.

Developed with the analyzer's semantic vocabulary, ConcurrBench establishes
conformance to intended distinctions, not general recall on unseen native code.

\find{\textbf{Answer to RQ1.} \tech{} correctly classifies all 180 cases.
Relative to dependency-only reporting, full state-aware analysis raises
precision from 50.85\% to 100\% while retaining 100\% recall. On this
analyzer-developed benchmark, the six-family results support the contract
distinctions, and the cumulative reductions support event filtering and
property-specific validation.}

\subsection{RQ2: Efficiency on CPython}\label{sec:rq2}

\noindent\textbf{Result.}
Andersen analysis has a 22.13--27.66-second median and 670--784\,MiB median
peak RSS across the three releases. The engine/plugins take 1.39--2.45 seconds;
the 3.14 flow-sensitive run reaches the 48\,GiB limit without a report.

\begin{table*}[!t]
\centering
\caption{Analysis time and memory by component.  Phase cells report seconds /
increase in peak resident set size (RSS), in MiB; end-to-end cells report
seconds / absolute peak RSS (MiB).  ``Killed'' means that no candidate report
was produced.}
\label{tab:performance}
\small
\begin{tabular}{@{}llrrrrl@{}}
\toprule
Target & Points-to & Frontend & Interproc. + \ncdg{} & Engine & Total & Outcome \\
\midrule
3.13 & Andersen & 13.57/305.25 & 7.17/306.56 & 1.39/0.15 & 22.13/670.1 & Complete \\
3.14 & Andersen & 15.33/354.64 & 9.35/357.91 & 2.45/0.23 & 27.00/770.7 & Complete \\
3.15 & Andersen & 16.64/354.39 & 9.12/370.22 & 1.90/0.17 & 27.66/783.7 & Complete \\
\midrule
3.14 & Flow-sensitive & -- & -- & -- & 413.16/49,054.3 & \makecell[l]{Memory limit;\\no report} \\
\bottomrule
\end{tabular}
\end{table*}

\noindent\textbf{Method.}
We run five complete analyses per release on the setup machine, sequentially with
rotating target order. Table~\ref{tab:performance} gives medians for frontend,
interprocedural/\ncdg{}, and engine/plugin phases. We measure wall time and each
phase's peak resident set size (RSS) increase; end-to-end RSS is the process's
absolute peak. The flow-sensitive baseline uses identical CPython 3.14 input and
options except points-to mode, with a 24-hour budget, 48\,GiB cgroup limit, and
no swap. Because the out-of-memory kill prevented profiler flushing, its RSS is
kernel-reported anonymous RSS at termination.

\noindent\textbf{End-to-end and component cost.}
Andersen runs~\cite{andersen1994} have ranges of 22.00--22.89,
26.81--28.03, and 27.37--28.22 seconds for 3.13, 3.14, and 3.15, with identical
candidate counts across repetitions. Frontend time dominates, followed by
interprocedural analysis/\ncdg{} construction. Engine/plugins take
1.39--2.45 seconds and add under 0.24\,MiB to the RSS high-water mark. Incremental
phase memory identifies when RSS grows, not all memory owned by that phase.

\noindent\textbf{Baseline comparison.}
The flow-sensitive run hits its 48\,GiB cgroup limit and is killed after
413.16 seconds without a report; the kernel records 47.90\,GiB anonymous RSS.
Andersen completes the same 3.14 input in 27.00 seconds (median) at
770.7\,MiB peak RSS. This comparison supports Andersen as the scalable default
for this environment. Accuracy remains unmeasured.

Analyzing CPython 3.15's 15,518 lifted functions and 510,614 normalized native
instructions in a 27.66-second median supports repeated release-scale offline
use.

\find{\textbf{Answer to RQ2.} Every release has an end-to-end median below 28 seconds and
0.8\,GiB median peak RSS; engine/plugins take at most 2.45 seconds. The
flow-sensitive baseline exhausts 48\,GiB without a report. This supports Andersen
as the scalable default on these release-scale inputs. The incomplete run leaves
accuracy unmeasured.}

% Float placement is not forced at the RQ boundary.

\subsection{RQ3: Real-Bug Detection and Confirmation}\label{sec:rq3}

\noindent\textbf{Results overview.}
Frozen reports contain 1,094 confirmed fingerprints (25.60\% yield), representing
139 qualified FT roots. The 144-root cross-mode inventory has strongest evidence of
125 native failures, 17 semantic or invariant violations, and two scheduling
observations backed by source-established invalidation intervals.
Tables~\ref{tab:root-accounting} and~\ref{tab:evidence-ledger} separate detector
attribution, root--version coverage, and root-level strongest evidence.

\noindent\textbf{Method.}
We analyze every compilation-database translation unit and review deduplicated
state-violation fingerprints. Source audit checks seven facts: real acquisition
and use; valid dependency; reachable event; relevant may-target with a concrete
match established for confirmed roots; required state transition; no covering
ownership, phase, or synchronization rule; and later use observing invalidity.
Confirmation requires this chain and dynamic or equivalent causal root evidence.

\noindent\textbf{Confirmation and PoC development.}
Each PoC follows its candidate's required transition.
Where possible, tests share only the implicated owner, perform invalidation in
one thread or callback, and force retained use in another operation. Oracles
check documented invariants where available; otherwise output supports scheduling
only, with causality grounded in native lifetime, sanitizer, or source evidence.
We inspect native stacks and use compatible ASan builds~\cite{asan}. Parallel
tests have same-version GIL or serialized controls; re-entry bugs are expected
to reproduce in both modes. A clean stress run covers one schedule. Source audit
then checks the candidate, and later tests focus on the exact transition.

Section~\ref{sec:impl}'s script proposes inheritance only for unchanged
source intervals, exact candidate semantics, and status-preserving normalized
event profiles. A reviewer separately approves the same causal native root and
checks conditional compilation and surrounding protection; changed acquisitions,
protection, or ambiguous matches require fresh audit. PoCs are replayed on target
releases where possible. The dataset retains roots, affected source,
manifestations, evidence, and runnable PoCs. Investigation-discovered acquisitions
absent from frozen candidates are initial detector gaps, excluded from frozen
yield and used only for retrospective catalog closure, leaving scored runs
unchanged.

\noindent\textbf{Focused qualification completion.}
For seven packages lacking current raw traces, 40 retained FT, GIL-control, and
retry runs yield 17 assertion/ASan aborts and 23 clean exits.
\texttt{OrderedDict} and writer-stats add no fresh failure; packages retain
prior positives and exact causal source intervals.
Appendix~\ref{app:qualification} provides per-package counts, protocol, and
intermittent SQLite results; commands and file records remain in the artifact. These runs
supplement the corpus ledger, not define its 144-root denominator.

\noindent\textbf{Production capability ablation.}
We freeze the analyzer, released catalogs, source snapshots, compilation
databases, Andersen settings, and root matchers, then rerun Full and three
one-capability variants. Direct relations retain only exact named-catalog
instantiations; local events exclude cross-entry connections; coarse protection
suppresses an event only when one guard covers all of its retained effects.
Conformance tests verify that each switch changes only its stated capability.
Three workers, one per release, run configurations sequentially.
The all-profile match measures evidence completeness: it requires every retained
confirming profile. The representation match measures detection: the same exact
acquisition/use fingerprint and property set retains at least one complete,
audited causal profile. Both metrics use semantic witnesses instead of
same-function warnings.

\begin{table*}[!t]
\centering
\caption{One-capability-at-a-time production ablations. Each cell gives
dependencies / fingerprints / all-profile matches / represented roots
(D/F/A/R).}
\label{tab:production-ablation}
\small
\setlength{\tabcolsep}{5pt}
\begin{tabular}{@{}lrrr@{}}
\toprule
Configuration & CPython 3.13 D/F/A/R & CPython 3.14 D/F/A/R & CPython 3.15 D/F/A/R \\
\midrule
Full & 5,674/1,546/115/115 & 6,061/1,459/105/105 & 6,240/1,502/90/92 \\
Direct relations only & 1,904/685/71/71 & 2,043/671/80/80 & 2,134/756/54/56 \\
Local events only & 5,674/293/24/27 & 6,061/288/8/11 & 6,240/353/18/19 \\
Coarse protection & 5,674/1,548/115/115 & 6,061/1,459/105/105 & 6,240/1,502/90/92 \\
\bottomrule
\end{tabular}
\end{table*}

Full recovers 1,904/2,043/2,134 direct catalog dependencies,
253/295/328 catalog-derived wrapper relations, and 3,517/3,723/3,778
source-inferred relations on 3.13/3.14/3.15. Direct-only mode contains direct
facts and loses 44, 25, and 36 represented roots. Local-only mode keeps the
Full dependency population but loses 88, 94, and 73 represented roots. Bound-64
checks recover none of these losses, distinguishing capability removal from
provenance truncation. Every ablation root also appears in Full. Across
variants, runs take 23.13--29.43 seconds
and 669.4--784.9\,MiB peak RSS under the same measurement procedure.

Full represents 115/121, 102/113, and 92/95 roots in the release-specific FT
inventory. The remaining ledger records identify twelve cases without a
historical fingerprint index and five 3.14 roots without an exact Full semantic
report. Three other 3.14 matches concern roots outside that release's FT
inventory and are outside the 102/113 count.

The complete semantic-difference ledger contains 6,371 records. Exact joins to
retained positive evidence support 1,586; the fixed-seed sample is drawn from
the other 4,785. Case-level audit reconstructs the dependency, event chain,
properties, and source intervals for all 182 sampled records. Sixty are
representation-only profile changes. Two coarse-only warnings are refuted by
their enclosing dictionary critical sections. The other 120 remain unresolved
with a causal blocker: target identity (40), API obligation (37), protection
interval (1), or insufficient causal evidence (42). Configuration was hidden,
but the packet exposed difference reason and side; we record that blinding
limitation.

\noindent\textbf{Bounded-witness diagnostic.}
The default limit of 16 discards unique external-entry
dependency/property/event provenance in Full, so we rerun affected
configurations at 64. Local-event provenance is deduplicated but unbounded.
CFG joins carry state alone; the dependency-level transition ledger retains
provenance.

\begin{table}[!htb]
\centering
\caption{Full-configuration witness-bound diagnostic. ``Affected'' counts
dependency/property diagnostic entries; $\Delta$ profiles counts fingerprints
whose retained event set changes.}
\label{tab:witness-bound}
\scriptsize
\setlength{\tabcolsep}{0.4pt}
\begin{tabular}{@{}lrrrr@{}}
\toprule
Release & Rejects 16$\rightarrow$64 & Affected 16$\rightarrow$64 & $\Delta$ profiles & A/R 16$\rightarrow$64 \\
\midrule
3.13 & 77,585$\rightarrow$54,919 & 876$\rightarrow$325 & 509 & 115/115$\rightarrow$115/115 \\
3.14 & 73,056$\rightarrow$50,853 & 851$\rightarrow$327 & 472 & 105/105$\rightarrow$105/105 \\
3.15 & 244,549$\rightarrow$220,068 & 818$\rightarrow$319 & 540 & 90/92$\rightarrow$92/92 \\
\bottomrule
\end{tabular}
\end{table}

Raising the bound leaves semantic warning sets unchanged. It expands retained event
references from 13,534/12,208/13,811 to 30,027/28,323/30,510 and supplies the
alternative profiles needed for two additional 3.15 all-profile matches:
\texttt{unicode-fromid-array-growth-race} and
\texttt{unicode-interned-size-unlocked-iteration}. Every discarded event
identifier remains in the serialized NCDG. Root matching credits report
witnesses; serialized graph presence supports audit. Represented-root sets stay
unchanged. Retained local
provenance reaches 208/221/227 events per dependency/property. The 64-bound Full reruns take
21.80--27.74 seconds and 672.9--782.4\,MiB. Thus, the larger provenance bound
keeps semantic reports and material cost stable in this batch.

\noindent\textbf{Protection-policy diagnostic.}
Full and coarse protection make different internal decisions only on 3.13:
22 partial-protection dependency/event pairs concentrated in two dependencies.
Both produce coarse-only warnings. Source audit refutes both because
\texttt{dict\_dict\_merge} holds a two-dictionary critical section and reverse
iteration holds the dictionary critical section across the complete use.
The policies never diverge on 3.14/3.15. Focused mixed-effect controls pass,
and the instrumented Full rerun reproduces every original Full fingerprint.

\noindent\textbf{Real vulnerable/repaired pair.}
We test an existing repair under a fixed semantic model. CPython 3.13's
one-argument \texttt{str.maketrans} traverses a caller-owned dictionary with
\texttt{PyDict\_Next}, consuming borrowed keys/values without the critical
section required under possible concurrent mutation
~\cite{py-free-threading-extensions}. L3 emits two acquisition candidates;
isolated key/value-mutation PoCs fail on pinned 3.13 FT ASan, while same-binary
GIL controls remain clean. Version 3.14 wraps traversal in
\texttt{Py\_BEGIN\_CRITICAL\_SECTION(x)} and both FT controls are clean.
A full 3.14 run with the unchanged 3.13 catalog emits no \texttt{maketrans}
candidate. This result is consistent with added interval protection under a
fixed catalog. Other source changes remain possible factors in the cross-release
comparison.

\noindent\textbf{Detector attribution and evidence provenance.}
Table~\ref{tab:root-accounting} separates frozen output from the study inventory:
139 of 141 qualified FT roots occur in frozen reports. The two setup acquisition
gaps belong to the inventory and are outside that detector run. Released runs
recover both and cover 141/141 qualified roots. Confirmation yield remains tied to
the frozen runs.

\begin{table*}[!t]
\centering
\caption{Authoritative detector and root accounting.  Released candidate
counts are unique within CPython 3.13/3.14/3.15; qualified-root coverage does
not estimate completeness beyond the study corpus.}
\label{tab:root-accounting}
\small
\setlength{\tabcolsep}{3pt}
\begin{tabular}{@{}p{.23\textwidth}p{.23\textwidth}p{.26\textwidth}p{.24\textwidth}@{}}
\toprule
Configuration or provenance & Candidate scope & Confirmation metric & FT-root attribution \\
\midrule
Frozen scored reports & 4,273 unique across releases & 1,094 confirmed (25.60\% yield) & 139/141 qualified roots represented \\
Investigation-discovered gaps & 2 missing acquisitions & excluded from scored yield & 2 roots \\
Released configuration & 1,546 / 1,459 / 1,502 & not re-adjudicated & 141/141 qualified-root coverage \\
\bottomrule
\end{tabular}
\end{table*}

A retrospective join of per-version inventories, evidence packages, and explicit
inheritance creates a 418-row root--version--runtime ledger. Rows identify
positive observations, artifact paths, source audits, applicable inheritance
sources, and runtime build/settings. A checked script reproducibly parses
authoritative tables, rejects denominator mismatches, and never treats file
presence or clean execution as positive evidence.

\begin{table}[!htb]
\centering
\caption{Auditable evidence inventory.  Strongest-evidence classes are
exclusive at root level; coverage counts use root--version rows.}
\label{tab:evidence-ledger}
\small
\setlength{\tabcolsep}{3pt}
\begin{tabularx}{\columnwidth}{@{}Xrl@{}}
\toprule
Ledger entry or evidence class & Count & Unit \\
\midrule
\multicolumn{3}{@{}l}{\emph{Root--version coverage}} \\
FT build, GIL disabled & 329 & root--version rows \\
Conventional-GIL build & 89 & root--version rows \\
Explicit equivalence inheritance & 242 & rows in both modes \\
\addlinespace
\multicolumn{3}{@{}l}{\emph{Strongest retained evidence per root}} \\
Native failure & 125 & unique roots \\
Semantic/invariant violation & 17 & unique roots \\
Scheduling observation + source proof & 2 & unique roots \\
\midrule
Unique study inventory & 144 & cross-mode roots \\
\bottomrule
\end{tabularx}
\end{table}

The 125 native-failure roots retain crashes, sanitizer errors, failed native
assertions, or invalid native accesses. Audit of the other 17 distinguishes one
documented public guarantee, four internal implementation invariants, and
twelve adopted correctness expectations. The \texttt{object.\_\_getstate\_\_}
case uses the absence of a legal old-or-new result under its controlled input as
an adopted correctness expectation, separate from a public guarantee. For \texttt{memoryview.tobytes} and noncontiguous
\texttt{memoryview.hex}, mixed contents establish only overlap; qualification
combines that observation with audited lease revocation. The evidence audit
checks all 144 roots and 418 root--version--runtime rows. It finds retained
observation and causal material for every root; all 418 rows have normalized
technical status and all 144 roots pass the category audit. Five formerly
missing rows used the equivalent \texttt{classification} field, while the
structseq package already retained direct failure evidence. A separate grouping audit resolves six apparent
multi-package merges as FT/conventional-GIL packages with the same root and
resolves every merge/split decision.
FT inheritance comprises 97 entries to 3.13, 90 to 3.15, and 142 version-local
or 3.14-baseline audits. Maintainer acknowledgment/fixes remain separate from
technical qualification.

\noindent\textbf{Prior public status.}
We search CPython issues and pull requests through September 6, 2026, before
the study's aggregate report. Search keys include native identifiers, source
paths, symptoms, and free-threading terms. A match covers the same native cause
or a specific matching item and fix in an umbrella report. Broad tracking
issues do not count by themselves. Study-authored records are excluded. The
audit maps 49 of 144 roots to 36 independent prior records. For the other 95,
we find no earlier public report of the same native root. We call these 95
previously unreported under this audit. The artifact ledger records every root,
record, and exclusion. This result covers public CPython history, not private
reports or sources outside the audit.

\noindent\textbf{Frozen confirmation yield.}
Table~\ref{tab:plugin-findings} reports emitted/confirmed fingerprints by
mechanism; the union deduplicates cross-release fingerprints. Overall yield is
1,094/4,273 (25.60\%); per-release yields are 31.07\% for 3.13 (480/1,545),
22.22\% for 3.14 (324/1,458), and 22.13\% for 3.15 (332/1,500).

RQ1 provides oracle-backed precision; production 25.60\% is dependency--event
confirmation yield. The remaining 3,179 fingerprints lack normalized
refuted/unresolved labels. Source review encountered private construction, shutdown-only
paths, immutable exact types, may-alias merging, and unreachable generic events
among non-confirmed reports.

\begin{table*}[!t]
\centering
\caption{Candidate/true-positive (C/TP) fingerprints by
invalidation-mechanism plugin. A TP is a confirmed manifestation; the union
deduplicates fingerprints shared across releases.}
\label{tab:plugin-findings}
\small
\begin{tabular}{@{}lrrrrr@{}}
\toprule
Mechanism plugin & 3.13 C/TP & 3.14 C/TP & 3.15 C/TP & Union C/TP & Union yield \\
\midrule
Borrowed-object lifetime & 576/224 & 556/164 & 547/160 & 1,632/529 & 32.41\% \\
Storage generation & 502/76 & 482/50 & 497/61 & 1,341/178 & 13.27\% \\
Container traversal & 5/3 & 5/3 & 6/4 & 16/10 & 62.50\% \\
View/lease lifetime & 58/14 & 61/14 & 102/14 & 210/42 & 20.00\% \\
Callback re-entry & 132/52 & 110/29 & 111/31 & 343/108 & 31.49\% \\
Reference-drop lifetime & 272/111 & 244/64 & 237/62 & 731/227 & 31.05\% \\
\midrule
\textbf{Fingerprint total} & \textbf{1,545/480} & \textbf{1,458/324} &
\textbf{1,500/332} & \textbf{4,273/1,094} & \textbf{25.60\%} \\
\bottomrule
\end{tabular}
\end{table*}

Borrowed-object lifetime contributes 529 confirmed manifestations, reference-drop
227, and storage generation 178. Traversal's highest yield (62.50\%) comprises
only ten confirmations from 16 candidates. Storage generation's lowest (13.27\%)
includes private, unpublished storage unreachable by generic resize events.
Plugins therefore require distinct stability contracts despite sharing an engine.

\noindent\textbf{Bug roots and security impact.}
The inventory contains 141 FT roots (provenance in
Table~\ref{tab:root-accounting}) and 144 across FT/conventional-GIL builds,
spanning 61 logical components and 75 C files: containers, memoryview, pickle,
codecs, import, garbage collection, compilation, SQLite, XML, SSL, and
\texttt{ctypes}. Appendix~\ref{app:components} summarizes component coverage;
Appendix~\ref{app:bugcases} traces security consequences.

Reproductions cause native crashes, sanitizer-detected use-after-free/wild writes,
container/compiler corruption, and impossible results. Fifty-four of 144 roots
have primary CWE-416 labels~\cite{cwe416}; others primarily expose
shared-resource races~\cite{cwe362}, time-of-check-to-time-of-use (TOCTOU)
failures~\cite{cwe367}, and resource-lifetime violations~\cite{cwe664}. Five
roots concern CPython 3.15's new remote-debugging component supporting sampling
profilers~\cite{py315-whatsnew}.

\begin{table*}[!t]
\centering
\caption{Trigger assumptions and directly observed evidence for representative
roots. Native-harness evidence validates a runtime interval within CPython's
extension model.}
\label{tab:threat-evidence}
\small
\setlength{\tabcolsep}{3pt}
\begin{tabular}{@{}p{.21\textwidth}p{.23\textwidth}p{.26\textwidth}p{.26\textwidth}@{}}
\toprule
Root & Required control & Contract under test & Direct observation \\
\midrule
\texttt{memoryview.tobytes} & Concurrent public Python calls & Active native copy keeps its backing lease & Release/resize overlaps a successful mixed-content copy; source audit establishes lease loss \\
Shared \texttt{pyexpat} parser & Shared parser; recursive handler for GIL mode & CPython serializes or rejects a second entry before reusing one Expat parser & ASan heap-guard overwrite in FT; null error-state crash under re-entry \\
Borrowed module \texttt{\_\_dir\_\_} & Public replacement in FT; conforming GIL-release test callable in GIL mode & Borrowed callable remains live until dispatch & ASan use-after-free in native call dispatch \\
Re-entrant \texttt{TextIOWrapper.tell} & User-defined codec callback & Borrowed snapshot bytes remain live across decoder calls & Outer call consumes allocator-reuse bytes introduced after re-entry \\
\bottomrule
\end{tabular}
\end{table*}

\noindent\textbf{Representative bug details.}
The running-example PoC uses a view over a 512-MiB bytearray: one thread calls
\texttt{tobytes}; another releases the view, clears the exporter, and refills it
with different bytes. The FT copy contains 4,096 original and 536,866,816
replacement bytes; the conventional-GIL control completes five clean rounds.
Mixed contents show that copying overlaps release/resize. They provide no
atomic-snapshot guarantee. Causal lifetime evidence is the audited sequence: \texttt{release}
drops the managed export, resize succeeds only afterward, and \texttt{tobytes}
continues through its pre-release descriptor. Output supplies scheduling evidence;
the audited lease sequence supplies the causal lifetime evidence.

In \texttt{bytearray.\_\_init\_\_}, locking the resize but not the later destination
write lets another resize revoke its pointer. Compiler metadata dictionaries
are sized and traversed independently, so mutation corrupts generated code.
In \texttt{OrderedDict}, comparison callbacks rebuild cached keys despite a
critical section; audit identifies stale cached-table use. Focused FT and
compatibility-GIL reruns are clean, but a separate retained PoC triggers ASan
use-after-free on all three conventional builds. Clean reruns are neither new
manifestations nor refutations.

\noindent\textbf{Release and runtime-mode behavior.}
Table~\ref{tab:roots-modes} shows FT roots declining from 121 in 3.13 to 113
in baseline 3.14~\cite{pep779} and 95 in 3.15. The latter comprises 90 inherited
and five new remote-debugging roots. This is consistent with audited increases
in object locking, owned-reference APIs, and atomic protocols; source changes
also remove exposure and merge manifestations.

\begin{table}[!htb]
\centering
\caption{Confirmed root counts by CPython build and runtime setting.}
\label{tab:roots-modes}
\small
\begin{tabular}{@{}lrr@{}}
\toprule
Version & Free-threaded & Conventional GIL \\
\midrule
3.13 & 121 & 30 \\
3.14 & 113 & 28 \\
3.15 & 95 & 31 \\
\midrule
Unique across releases & 141 & 34 \\
\bottomrule
\end{tabular}
\end{table}

All 34 GIL roots have positive observations on separately compiled
conventional-GIL builds; four also reproduce with the FT binary's compatibility
GIL. The other compatibility runs serve as controls and are excluded from that
count. GIL-aware PoCs target
same-thread comparison, descriptor, or I/O callbacks, finalizers/signals, and
native blocking operations releasing the GIL. Static candidates come from FT
compilation databases; each GIL package audits
its conditional source path and qualifies a separately compiled interpreter.
The 141 FT and 34 GIL sets yield 144 roots: 110 FT-only, 31 shared, and three
GIL-only.

\begin{table*}[!t]
\centering
\caption{External baseline outcomes at each tool's stated input scope.  C
denotes a completed native result, TO a 24-hour timeout, FE a frontend or
unsupported-feature failure, and UA an unavailable licensed executable.  A
ConcurDep C cell gives raw/unique candidates; an SVF C cell gives race pairs.
TO, FE, and UA have no detection count.}
\label{tab:external-baselines}
\small
\begin{tabular}{@{}lllll@{}}
\toprule
Tool & Input scope & CPython 3.13 & CPython 3.14 & CPython 3.15 \\
\midrule
\tech{} & Complete compilation database & C: 1,564/1,546 & C: 1,473/1,459 & C: 1,514/1,502 \\
Goblint 2.8.0 & Whole-program accepted input & TO & TO & TO \\
RacerF 2.2.1 & Whole-program accepted input & FE & FE & FE \\
SVF-MTA/SlicedMTA & Complete LLVM bitcode set & C: 0 pairs & C: 0 pairs & C: 0 pairs \\
Deagle 4.1.0 & Thread-entry applicability pilot & FE & FE & FE \\
O2 1.1.3 & Licensed executable & UA & UA & UA \\
\bottomrule
\end{tabular}
\end{table*}

\noindent\textbf{External baseline applicability.}
Table~\ref{tab:external-baselines} covers the same pinned releases. Goblint
accepts 281/284, 297/303, and 301/309 units but times out at 24 hours
~\cite{goblint}. RacerF~\cite{racerf} accepts 271/284, 275/303, and 277/309
before Frama-C rejects conflicting \texttt{pthread\_sigmask} declarations.
A declaration-only 3.13 header repair progresses to incompatible
\texttt{syscall} declarations, still FE. Deagle~\cite{deagle} reaches GOTO
generation for \texttt{PyThread\_start\_new\_thread} but rejects
\texttt{Include/object.h}'s thread-pointer assembly. Replacing it would change
thread identity, so the pilot remains FE.

SVF MTA~\cite{svf,sui2016sparse,svf-slicedmta} completes full bitcode inputs
(284/284, 303/303, 309/309 units) with zero SlicedMTA race pairs. The same
installation detects a pthread write/write positive control; a diagnostic 3.13
run recognizes \texttt{pthread\_create} and constructs worker-thread nodes.
These are zero-result runs under SVF's model. They provide no race-freedom
claim. The tools also use different reporting units: SVF reports access pairs,
while \tech{} reports dependency-state violations. Goblint measures scalability, RacerF/Deagle applicability, and
O2~\cite{o2} remains UA without a licensed executable.

\find{\textbf{Answer to RQ3.} Frozen reports confirm 1,094/4,273 fingerprints
(25.60\% yield), representing 139 FT roots. Released catalogs repair two gaps
and cover all 141 qualified FT roots. The 144-root cross-mode inventory includes 125
native failures; all 34 GIL roots have conventional-build evidence. Direct-only
relation recovery loses 25--44 represented roots per release, and local-only
events lose 73--94; coarse protection loses none. A larger external-witness
bound changes no semantic warning or represented root, but completes two
additional 3.15 all-profile matches. The
external matrix measures tool applicability.}

\section{Discussion}\label{sec:diss}

\noindent\textbf{\circleone{1} Why Dependency Invalidation Helps.}
Race reports identify concurrent accesses lacking happens-before order;
\tech{} identifies state a later native use assumes and the event revoking it.
This explains failures across fields or allocations: dictionary deletion versus later
increment of its former value, bytearray resize versus a previously returned
pointer, or view closure versus a later lease-dependent copy. Dependency failure
need not involve same-address instructions.

The abstraction unifies parallel and re-entrant failures: in
Appendix~\ref{app:bugcases}'s \texttt{TextIOWrapper.tell}, a decoder callback
releases a borrowed snapshot on the same thread despite the GIL. Both schedules
ask whether an event within the dependency interval invalidates its contract.
This case exercises different information from the view lease. Dependency
recovery links borrowed \texttt{next\_input} to \texttt{self->snapshot}. The
NCDG follows the decoder summary to re-entrant \texttt{seek}, which clears that
exact snapshot. Validation recognizes that the object critical section excludes
a peer thread while allowing the same-thread callback. Per-object liveness or
a same-address race alone would lose this owner relation and re-entry chain.
The abstraction may extend to native reference, view, handle, or iterator APIs;
this evaluation covers CPython.

\noindent\textbf{\circleone{2} Confirmation Yield and Specification Cost.}
At 25.60\% yield, production reports need source review. Non-confirmed reports
often lack owner facts: pre-publication privacy, unexposed compiler construction,
teardown exclusion, or different generic-container instances. More precise
escape, phase, and relational heap abstractions could remove them.

Catalogs make runtime knowledge auditable and plugins independent of CPython
names, but omissions can cause false negatives, and release changes need maintenance.
The frozen catalogs contain 338--397 facts. Setup construction/coverage audits
define validated inputs; post-audit catalogs recover both known omissions.
Unknown idioms remain outside measured scope. Source-equivalent inheritance
reduces re-review; extraction from C annotations and generated Clinic metadata
is promising future work.

The production capability ablation separates supplied and inferred facts.
Direct catalog rules account for 1,904--2,134 dependencies; wrapper derivation
and source inference add 3,770--4,106 and retain 25--44 further represented
roots. Cross-entry event connection retains 73--94 represented roots relative
to local-only analysis. Coarse protection keeps the represented-root set. Its only
internal divergence is 22 decisions over two 3.13 dependencies; both coarse-only
warnings are source-refuted because object critical sections cover their uses.

One root investigation may resolve several fingerprints. Without prospective
review-time or refuted/unresolved labels, we report audit units and yield, not
person-hours or counted false-positive causes.

\noindent\textbf{\circleone{3} Limitations.}

\noindent\textbf{Design limitations.}
\tech{}'s context-insensitive points-to analysis can merge unrelated heap objects;
target abstraction can merge shared types/field paths. CFG/interprocedural
reachability approximates branch feasibility, retaining potential cross-entry
invalidations at the cost of false positives. Missing contracts, acquisition
patterns, or state dimensions cause false negatives. Plugins are modular within
existing state vocabulary; new properties
or effects need core lattice and transfer extensions. Reports identify possible
invalidations, leaving schedule synthesis and exploitability downstream. The
memory model covers cataloged atomics, locks, lifetimes, and exposure; unmodeled
fences, lock-free reclamation, signals, or re-entrant critical sections may escape
it. Single external-event alternatives and exclusion of owned/pinned borrows
from re-entry are further false-negative boundaries. The latter comes from the
selector. Storage analysis can report resize despite owner liveness. The
re-entry selector removes relation-changing callbacks on owned borrowed-object
dependencies before checking relation state. Promotion
checks and discharges only the borrowed-object obligation; separate storage/lease
dependencies remain. Even small release changes require inheritance review.

\noindent\textbf{Implementation limitations.}
Clang~14 analyzes the three trees' Linux FT compilation-database units. Inline
assembly, excluded platform modules, unmodeled libraries, and uncataloged
extensions remain opaque. Catalogs and ConcurrBench were co-developed;
third-party extensions and other runtimes remain future evaluation. The five
new 3.15 remote-debugging roots are not held-out recall evidence because catalog
chronology was not prospectively frozen. Catalog construction, root
deduplication, and final audit are manual. The optional profiler uses Linux
\texttt{/proc}; the analyzer is otherwise platform-independent at that
interface. PoCs are developed separately; the prototype synthesizes neither
schedules, tests, nor patches. Conventional-GIL reasoning requires cataloged
callback, finalizer, signal, foreign-thread, and GIL-release mechanisms.

\noindent\textbf{\circleone{4} Validation and Threats to Validity.}

\noindent\textbf{Ground truth.}
ConcurrBench tests implementation fidelity; CPython adds real-world diversity
but relies on manual labels/root grouping. Validation uses target-specific
negative controls, same-object tests, native stacks, and root deduplication.
Clean stress leaves source-valid risks open; crashes sharing a root count once.
Independent source passes cover a subset. Without full-dataset reviewer
agreement, grouping, primary-CWE labels, and inheritance remain judgment-sensitive.
The retrospective evidence audit checks every root record. All 418 rows have a
normalized technical status, and all 144 roots pass the evidence-category
audit. The semantic audit further separates public guarantees, internal
invariants, and study-adopted correctness expectations.

\noindent\textbf{Catalog completeness.}
Released 141/141 coverage in Table~\ref{tab:root-accounting} measures the
qualified FT root corpus. ConcurrBench supplies the recall denominator.
Unrecognized idioms can still cause misses.

\noindent\textbf{Version and platform scope.}
The three Linux/x86-64 snapshots span experimental 3.13 FT
~\cite{py313-free-threading}, supported baseline 3.14~\cite{pep779}, and
release-candidate 3.15. Compilers, allocators, architectures, and later source
changes affect manifestation. Equivalent-source inheritance broadens longitudinal
coverage, but future versions still need testing. Scheduling, allocator behavior,
assertions, and sanitizer availability also affect PoCs; clean execution supports
only that run.

\noindent\textbf{Comparative scope.}
Cumulative ConcurrBench ablations measure exposure, target/interval, and
state/protection contributions. Production capability ablations report both
all-profile retention and one-complete-witness root representation. Among a
fixed sample from 4,785 previously unmatched differences, 60 are
representation-only, two are refuted, and 120 retain explicit causal blockers.
The sample records dispositions without estimating precision. The external
matrix measures acceptance and completion of stated inputs. An accuracy
comparison between SVF access pairs and \tech{} dependency-state violations
requires a shared oracle. Dynamic race and memory tools support qualification
through different reporting units.

\noindent\textbf{\circleone{5} Security Impact and Ethics.}

The inventory's concrete security failures include native crashes,
use-after-free/wild writes, traversal invalidation, and corrupted runtime data
through public Python/C-API paths. They threaten availability and memory
integrity; use-after-free can enable crashes, corruption, and code
execution~\cite{cwe416}. Reproduction establishes technical impact, while
practical exploitability depends on API exposure, allocator behavior, privileges,
and the embedding application.

High/Critical labels prioritize mechanism-family review, not deployment severity.
Each root receives one primary CWE label. The 54 CWE-416 roots count primary
labels; other roots can also include use-after-free manifestations.

Experiments use local CPython builds and synthetic inputs, without third-party
services, user data, or unsolicited scanning. Artifacts separate evidence,
versions, and modes. The study filed an aggregate CPython report on September
16, 2026, covering the then-current 142-root inventory. It is study disclosure,
so the prior-public audit excludes it. As of September 22, the disclosure ledger
attributes no maintainer acknowledgment, CVE, or fix to that report. The two
later catalog-gap roots remain for follow-up. Earlier independent issues and
patches provide technical context, not study-disclosure outcomes.

\section{Related Work}\label{sec:related}

\noindent\textbf{\circleone{1} Dynamic Race and Memory-Error Detection.}
Dynamic race detectors infer missing synchronization from executions.
ThreadSanitizer combines happens-before, locksets, and annotations for large
C/C++ systems~\cite{tsan}; AddressSanitizer instruments accesses and poisons
freed/out-of-bounds regions to identify manifested errors~\cite{asan}. Both
complement \tech{}: sanitizers qualify findings, while static reasoning connects
runtime acquisitions, invalidating events, and later uses. Dynamic tools need
triggering inputs/schedules; \tech{} trades that requirement for false positives.

\noindent\textbf{\circleone{2} Static Data-Race Analysis.}
Static race analyses combine aliasing, locksets, escape reasoning, and summaries.
Locksmith correlates C locks, shared locations, and accesses~\cite{locksmith};
Goblint uses thread-modular abstract interpretation with lock/access reasoning
~\cite{goblint}. Chord applies context sensitivity and staged Java pruning
~\cite{chord}; RacerD uses scalable, low-cost compositional summaries~\cite{racerd}.
O2 unifies thread/event entry contexts through origins~\cite{o2}. Goblint and
RacerD represent thread-modular and compositional approaches. \tech{} also uses
points-to facts, target matching, and summaries but checks acquisition-to-use
dependencies whose validity events change, spanning locations and runtime
ownership or buffer-lease relations.

Clang Thread Safety Analysis checks annotated locking disciplines
(\texttt{GUARDED\_BY}, \texttt{REQUIRES}, \texttt{EXCLUDES})
~\cite{clang-thread-safety}, complementing \tech{}'s recovery from unannotated
implementation code.

\noindent\textbf{\circleone{3} Python/C Lifecycle and Interference-Aware Typestate.}
CPyChecker checks extension-function reference counts path-sensitively using
modeled borrows/stolen references~\cite{cpychecker}. Pungi transforms Python/C
interface code into affine programs for interprocedural reference-count checking
~\cite{pungi}. PyRefcon symbolically tracks lifecycle/refcount transitions
through Python/C APIs and pointers in its PyObject State Transition
Model~\cite{pyrefcon}. Pungi and PyRefcon directly address native object lifetime,
closer to \tech{} than generic race detection, but primarily track individual
PyObject states/counts. \tech{} recovers owner--subject--storage relations and
parallel/re-entrant revocation of containment, address, traversal, or lease
obligations. This relational, cross-entry scope covers live owners with changed
subjects or backing relations.

Typestate encodes legal sequences in state-sensitive types~\cite{strom1986typestate};
ownership types constrain aliasing/shared access~\cite{clarke1998ownership}.
Typestate-guided concurrency combines transitions and inferred locksets for
shared-variable interference~\cite{typestate-concurrency}. \tech{} shares this
state-change principle, specializing in recovery of implicit CPython relations
from native operations/catalogs, matching cross-entry events, and applying
property-specific ownership, lease, revalidation, and protection. It checks
existing unannotated code under Section~\ref{sec:diss}'s catalog and may-analysis
limits.

JATO enforces atomicity of Java native methods invoked through the Java Native
Interface (JNI)~\cite{jato}. \tech{} instead checks property-specific relational
invalidation by peer-entry and callback events inside acquisition-to-use
intervals.

\noindent\textbf{\circleone{4} Testing Language Runtimes.}
Runtime fuzzers expose interpreter/cross-language failures using high-level
programs and native interactions. PyRTFuzz combines Python/native fuzzing
~\cite{li2023pyrtfuzz}; UAFL guides fuzzing toward use-after-free with static
typestate sequences~\cite{uafl}. Fuzzing produces concrete failures along the
semantic setups and schedules it explores. \tech{} reports source-derived
dependency/event chains for downstream dynamic confirmation. Schedule-aware test
generation is a natural extension.

\section{Conclusion}\label{sec:conclude}

Free threading exposes native safety's dependence on ownership, storage,
traversal, lease, version, and callback-sensitive relations absent from C types.
\tech{} treats each as a dependency whose required properties a target-matched
event may revoke before use. Recovery, event-aware \ncdg{} links to parallel and
re-entrant events, and property-specific state/protection validation implement
this insight. All 180 controlled cases are classified correctly; each selected release
completes below 28 seconds and 0.8\,GiB peak RSS in five-run medians. Frozen
reports represent 139 FT roots; two repaired catalog gaps yield all 141 qualified
FT roots in the study corpus. The study qualifies 144 roots across
FT/conventional-GIL builds,
including 125 with native-failure evidence. A public-history audit finds no
earlier report of the same native root for 95 of them. Production ablations show that
derived relation recovery and cross-entry event connection retain represented-root
evidence that restricted variants lose; witness-bound reruns preserve the
semantic warning set. These results establish dependency invalidation as a
practical unit for failures crossing
references, representations, callbacks, and execution modes.

{\small
\bibliographystyle{IEEEtran}
\bibliography{reference}
}

\appendices

\refstepcounter{section}
\section*{Appendix~\thesection. Plugin Specifications}
\label{app:plugins}

Section~\ref{sec:runtime-state} defines the common plugin interface and gives
the view/lease plugin in full.  This appendix specifies the other five
plugins.  For a transition set $\Theta$, all five use the same acceptance
predicate
\[
\mathsf{Qual}(\Theta)\equiv\exists\theta\in\Theta.\;
\theta.o\land\theta.f\land\theta.t\land
\theta.\Delta\neq\emptyset\land\theta.V\neq\emptyset,
\]
where $o$, $f$, and $t$ denote temporal overlap, an eligible may-execute
context, and may-target matching.  The core computes these facts; a plugin selects only its applicable
dependencies and required stability properties.  We abbreviate the properties
from Table~\ref{tab:contracts} as \textsc{Addr}, \textsc{Repr},
\textsc{Struct}, \textsc{Iter}, \textsc{Lease}, \textsc{Open},
\textsc{Callback}, \textsc{Rel}, and \textsc{Live}.

\noindent\textbf{\circleone{1} Storage-generation invalidation.}
For an interior address, the address and its representation remain valid until
the last dereference:
\[
\begin{aligned}
Q_S(D,G)&\equiv D.h=\mathsf{InteriorStorage},\\
K_S(D)&=\{\mathsf{Addr},\mathsf{Repr}\},\\
A_S(D,\Theta,G)&\equiv\mathsf{Qual}(\Theta_S).
\end{aligned}
\]
The plugin assigns high review priority and reports the resize, reallocation, rebuild,
close, or free that can revoke the address.

\noindent\textbf{\circleone{2} Container-traversal invalidation.}
For a size, index, cursor, or entry produced while traversing a container,
\[
\begin{aligned}
Q_T(D,G)&\equiv D.h=\mathsf{ContainerTraversal},\\
K_T(D)&=\{\mathsf{Struct},\mathsf{Iter}\},\\
A_T(D,\Theta,G)&\equiv\mathsf{Qual}(\Theta_T).
\end{aligned}
\]
The high-priority rationale identifies the structural mutation that can make
the saved bound, position, or entry relation stale.

\noindent\textbf{\circleone{3} Borrowed-object lifetime invalidation.}
Let $\mathcal{O}_{B}$ contain unowned and unpinned acquisition states. The
evaluated implementation conservatively requires both lifetime
and relation stability until promotion or final use:
\[
\begin{aligned}
\mathcal{O}_{B}&=\{\mathsf{borrowed},\mathsf{unowned},\mathsf{unknown}\},\\
Q_B(D,G)&\equiv D.h=\mathsf{BorrowedObject}
                 \land D.w\in\mathcal{O}_{B}\\
        &\quad\land D.U\neq\emptyset,\\
K_B(D)&=\{\mathsf{Live},\mathsf{Rel}\},\\
A_B(D,\Theta,G)&\equiv\mathsf{Qual}(\Theta_B).
\end{aligned}
\]
The rationale identifies the removal, replacement, or reference drop that may
invalidate subject lifetime or the acquisition relation. Here, \textsc{Rel}
means that the relation recorded at acquisition still holds. A report based
only on \textsc{Rel} enters contract audit. A separately pinned subject under a
snapshot-permitting contract requires \textsc{Live}, while membership may change.
A use that assumes the saved subject still occupies the owner's selected role
requires both properties. This conservative default contributes false positives
when recovery cannot distinguish those use contracts.

\noindent\textbf{\circleone{4} Callback re-entry invalidation.}
Let $\mathsf{ReentryEvidence}(D,G)$ denote a linked Python callback, native
callback, or event whose effect permits re-entry and mutation.  Owned or pinned
borrowed-object subjects are excluded.  Let
$B=\mathsf{BorrowedObject}$:
\[
\begin{aligned}
\mathcal{O}_{R}&=\{\mathsf{Owned},\mathsf{Pinned}\},\\
Q_R(D,G)&\equiv\mathsf{ReentryEvidence}(D,G)\\
 &\quad\land\neg(D.h=B
          \land D.w\in\mathcal{O}_{R}),\\
K_R(D)&=\{\mathsf{Callback},\mathsf{Rel}\},\\
A_R(D,\Theta,G)&\equiv\mathsf{Qual}(\Theta_R).
\end{aligned}
\]
The high-priority report names the callback that can replace or clear retained
state before native execution resumes.

This selector reflects the evaluated implementation: it excludes an owned or
pinned borrowed-object dependency from the re-entry plugin.  Ownership protects
lifetime but may leave the owner--subject relation unprotected. Relation-changing
re-entry for that excluded combination is a known false-negative boundary.

% Removed unconditional fill/break; keep the plugin in normal column flow.
\noindent\textbf{\circleone{5} Reference-drop lifetime invalidation.}
Let $\mathsf{FinalizationEvidence}(D,G)$ hold for a may-finalize effect,
finalization exposure, reference drop, or ownership transfer.  Then
\[
\begin{aligned}
Q_F(D,G)&\equiv\mathsf{FinalizationEvidence}(D,G),\\
K_F(D)&=\{\mathsf{Live}\},\\
A_F(D,\Theta,G)&\equiv\mathsf{Qual}(\Theta_F).
\end{aligned}
\]
This plugin receives the highest review priority because a drop-to-zero transition can execute
destructors and reclaim the exact subject before its final use.

The core supplies reachability and alias evidence to every predicate.  New
plugins therefore inherit the same target, overlap, and protection requirements
as existing families.

\refstepcounter{section}
\section*{Appendix~\thesection. NCDG Scale}
\label{app:ncdg-scale}

Table~\ref{tab:ncdg-scale} reports the graph objects serialized by each
whole-program run.  Event relations connect events to targets and contexts;
dependency--event links retain the candidate pairs considered by the state
engine after target indexing.  These counts complement the time and memory
results in Table~\ref{tab:performance}.

\begin{table}[!htb]
\centering
\caption{Scale of the event-aware NCDG for each CPython release.}
\label{tab:ncdg-scale}
\scriptsize
\begin{tabular}{@{}lrrrr@{}}
\toprule
\multicolumn{5}{@{}l}{\emph{Nodes and semantic objects}} \\
Version & Functions & Contexts & Dependencies & Events \\
\midrule
3.13 & 14,191 & 20,329 & 5,674 & 92,369 \\
3.14 & 15,356 & 21,594 & 6,061 & 104,282 \\
3.15 & 15,518 & 21,880 & 6,240 & 102,305 \\
\midrule
\multicolumn{5}{@{}l}{\emph{Graph relations and protection regions}} \\
Version & Call edges & Event rels. & Dep.--event links & Protection \\
\midrule
3.13 & 58,277 & 446,789 & 217,021 & 417 \\
3.14 & 64,573 & 503,981 & 528,380 & 784 \\
3.15 & 64,187 & 507,840 & 351,017 & 809 \\
\bottomrule
\end{tabular}
\end{table}

\refstepcounter{section}
\section*{Appendix~\thesection. Focused Qualification Runs}
\label{app:qualification}

We built the three pinned revisions with GCC~13, debug assertions,
AddressSanitizer, and free threading. A three-worker harness retained each
command, interpreter and PoC identifier, exit signal, standard output, and standard
error.  Table~\ref{tab:qualification-completion} reports the 30 primary jobs
and ten focused retries.  The SQLite and writer-stats rows include the primary
run plus five isolated free-threaded retries.

\begin{table}[!htb]
\centering
\caption{Focused dynamic qualification.  FT and GIL cells report
aborting/retained runs.}
\label{tab:qualification-completion}
\scriptsize
\begin{tabular}{@{}p{.38\columnwidth}rrp{.28\columnwidth}@{}}
\toprule
Root or mode group & FT & GIL & Outcome \\
\midrule
\texttt{partial} keyword repr & 1/1 & 0/1 & ASan crash during callback use \\
Import traceback chain & 1/1 & 0/1 & Debug invariant violation \\
\texttt{OrderedDict} cached keys & 0/1 & 0/1 & No fresh failure \\
SQLite blob registry & 4/6 & 0/1 & Intermittent free-threaded reproduction \\
Writer stats/close & 0/6 & 0/1 & No fresh failure \\
Writer internal state (seven modes) & 7/7 & 1/7 & GIL abort is \texttt{finalize\_close} \\
Writer input lists (three modes) & 3/3 & 0/3 & All free-threaded modes reproduce \\
\midrule
\textbf{Total} & \textbf{16/25} & \textbf{1/15} & \textbf{17/40 abort} \\
\bottomrule
\end{tabular}
\end{table}

A clean stress run covers only the exercised schedule.  SQLite illustrates
the converse: its primary run exits cleanly, whereas four of five isolated
retries abort.  The \texttt{OrderedDict} package retains conventional-GIL ASan
failures from all three pinned releases even though this focused rerun is
clean; writer-stats retains a prior free-threaded SIGSEGV description while all
six current free-threaded executions are clean.

\refstepcounter{section}
\section*{Appendix~\thesection. Aggregate Component-Coverage Summary}
\label{app:components}

Table~\ref{tab:component-summary} summarizes all 61 logical components that
contain a confirmed root.  Each version has separate free-threaded (FT) and
conventional-GIL counts.  The \emph{Unique roots} column deduplicates findings
across all releases and both modes.  We consolidate the library-level
\texttt{collections} entry and its \texttt{OrderedDict} implementation into one
logical component.

\begin{table}[!htb]
\centering
\caption{Root and component coverage by subsystem.  Version cells report
FT/conventional-GIL roots.}
\label{tab:component-summary}
\scriptsize
\setlength{\tabcolsep}{2pt}
\begin{tabular}{@{}lrrrrr@{}}
\toprule
Category & Components & 3.13 & 3.14 & 3.15 & Unique \\
\midrule
Library/extension & 29 & 61/22 & 62/19 & 49/22 & 78 \\
Built-in/core & 17 & 36/3 & 29/4 & 26/4 & 39 \\
Runtime/compiler & 15 & 24/5 & 22/5 & 20/5 & 27 \\
\midrule
\textbf{Total} & \textbf{61} & \textbf{121/30} & \textbf{113/28} &
\textbf{95/31} & \textbf{144} \\
\bottomrule
\end{tabular}
\end{table}

% IEEEtran places the appendix label and section title on separate lines for a
% numbered appendix.  Advance the counter explicitly and use one compact
% heading so the label and descriptive title remain together.
\refstepcounter{section}
\section*{Appendix~\thesection. Confirmed Bug Case Studies}
\label{app:bugcases}

The following cases use the same structure: the vulnerable source operation,
the invalidating event, the witness retained by \tech{}, the confirming PoC,
and the security consequence.  Together they cover the four primary CWE
classes used in the released bug dataset.

\par\medskip
\noindent\textbf{\circleone{1} CWE-362: shared \texttt{pyexpat} parser
\cite{cwe362}.}

\begin{tcolorbox}[colback=black!3,colframe=black!30,boxrule=0.3pt,arc=0.5mm,
  width=0.98\linewidth,left=0.8mm,right=0.8mm,top=0.5mm,bottom=0.5mm]
\scriptsize
\textbf{Root:} \texttt{pyexpat-shared-parser-native-state-race}\\
\textbf{Component:} \texttt{Modules/pyexpat.c} and Expat
\hfill \textbf{Affected:} CPython 3.13--3.15, FT and GIL\\
\textbf{Observed impact:} native heap corruption and process abort
\end{tcolorbox}

\noindent\hfill\begin{minipage}{0.96\linewidth}
{\scriptsize\textit{Unserialized parse entry:}}\par
\begin{lstlisting}[style=casecode,firstnumber=910]
while (slen > MAX_CHUNK_SIZE) {
    rc = XML_Parse(self->itself, s, MAX_CHUNK_SIZE, 0);
    ...
}
rc = XML_Parse(self->itself, s, (int)slen, isfinal);
\end{lstlisting}
\end{minipage}\hfill\mbox{}\par

\noindent\textbf{Trigger.}
Expat permits one active entry through each \texttt{XML\_Parser}; its API
specifically forbids calling parsing functions
on the active parser from one of its handlers~\cite{expat-api}.  The
Python-facing parser owns that single native handle. Its wrapper contract
serializes a second \texttt{Parse()} call or rejects it before entering Expat.
The free-threaded module instead allows
two public calls to enter Expat with the same handle.  Both mutate DTD tables, tag buffers,
entity lists, and parser position.  One call can therefore reallocate or free
state retained by the other.  \tech{} links the shared parser dependency to a
may-target second parse entry with an overlap context.  Source audit verifies
that both calls use the same parser and finds no parser-level critical section
preserving the native state.

\noindent\textbf{Validation and security impact.}
A barrier releases two \texttt{Parse()} calls on one parser.  The 3.14
free-threaded ASan build immediately overwrites the trailing guard of a 1 MiB
Expat allocation and aborts.  A separate conventional-GIL PoC recursively
calls \texttt{Parse()} from a handler; the nested call corrupts the outer
error state and passes a null error string into \texttt{strlen}.  The root also
reproduces on 3.15.  Untrusted XML supplied to a service that shares or
recursively enters a parser can thus cause native corruption and denial of
service in either runtime mode.

\par\medskip
\noindent\textbf{\circleone{2} CWE-367: stale inline-dictionary check
\cite{cwe367}.}

\begin{tcolorbox}[colback=black!3,colframe=black!30,boxrule=0.3pt,arc=0.5mm,
  width=0.98\linewidth,left=0.8mm,right=0.8mm,top=0.5mm,bottom=0.5mm]
\scriptsize
\textbf{Root:} \texttt{object-getstate-inline-dict-empty-transition}\\
\textbf{Component:} \texttt{Objects/dictobject.c}
\hfill \textbf{Affected:} CPython 3.13--3.15 FT\\
\textbf{Observed impact:} impossible empty serialization state
\end{tcolorbox}

\noindent\hfill\begin{minipage}{0.96\linewidth}
{\scriptsize\textit{Check and later slot scan:}}\par
\begin{lstlisting}[style=casecode,firstnumber=7657]
if (FT_ATOMIC_LOAD_UINT8(values->valid)) {
    PyDictKeysObject *keys = CACHED_KEYS(tp);
    PyObject **slots = values->values;
    for (...) {
        if (FT_ATOMIC_LOAD_PTR_RELAXED(slots[i]) != NULL)
            return 0;
    }
    return 1;
}
\end{lstlisting}
\end{minipage}\hfill\mbox{}\par

\noindent\textbf{Trigger.}
The reader checks \texttt{values->valid} only once.  A peer then assigns a new
\texttt{obj.\_\_dict\_\_}, publishing a nonempty dictionary while invalidating
and clearing the old inline slots.  The reader scans those cleared slots and
returns ``empty'' without rechecking the representation.  \tech{} records the
validity check, slot-dependent use, representation-changing store, and absent
revalidation as one state-transition witness.

\noindent\textbf{Validation and security impact.}
The PoC keeps both the old and replacement dictionaries nonempty, so
\texttt{object.\_\_getstate\_\_()} returning \texttt{None} has no legal
linearization point.  It does so after 3,182 reads on 3.15 FT; the same binary
with the GIL completes 50,000 rounds.  Serialization, checkpointing, or policy
code may silently omit security-relevant object state instead of raising an
exception.

\par\medskip
\noindent\textbf{\circleone{3} CWE-416: borrowed module
\texttt{\_\_dir\_\_}~\cite{cwe416}.}

\begin{tcolorbox}[colback=black!3,colframe=black!30,boxrule=0.3pt,arc=0.5mm,
  width=0.98\linewidth,left=0.8mm,right=0.8mm,top=0.5mm,bottom=0.5mm]
\scriptsize
\textbf{Root:} \texttt{module-dir-borrow-promotion}\\
\textbf{Component:} \texttt{Objects/moduleobject.c}
\hfill \textbf{Affected:} CPython 3.13--3.15, FT and GIL\\
\textbf{Observed impact:} use-after-free in native call dispatch
\end{tcolorbox}

\noindent\hfill\begin{minipage}{0.96\linewidth}
{\scriptsize\textit{Borrow followed by callback use:}}\par
\begin{lstlisting}[style=casecode,firstnumber=1193]
PyObject *dirfunc =
    PyDict_GetItemWithError(dict, &_Py_ID(__dir__));
if (dirfunc) {
    result = _PyObject_CallNoArgs(dirfunc);
}
\end{lstlisting}
\end{minipage}\hfill\mbox{}\par

\noindent\textbf{Trigger.}
Owning the module dictionary leaves a removed object's lifetime unprotected.
\texttt{PyDict\_GetItemWithError} returns a borrowed callable, and a peer can
replace \texttt{module.\_\_dir\_\_} after lookup but before the call; the C API
documents both the borrowed result and its need for external synchronization
under concurrent mutation~\cite{py-dict-api}.  The old
callable may reach zero references, while the reader dispatches through its
freed object and type.  \tech{} identifies the borrowed ownership, exact
dictionary replacement, drop-to-zero transition, and final call use.

\noindent\textbf{Validation and security impact.}
Eight \texttt{dir(module)} workers race one replacement thread.  The 3.14
free-threaded ASan build crashes in under one second from
\texttt{\_PyObject\_CallNoArgs}.  For the conventional-GIL build, a minimal C
callable releases the GIL inside \texttt{tp\_call}; a peer deletes the exact
dictionary entry, and ASan reports the same use-after-free when the call
resumes.  A native lifetime error in generic module introspection creates a
process-crash primitive and may permit stronger memory corruption under
favorable allocator reuse.

\par\medskip
\noindent\textbf{\circleone{4} CWE-664: reentrant
\texttt{TextIOWrapper.tell}~\cite{cwe664}.}

\begin{tcolorbox}[colback=black!3,colframe=black!30,boxrule=0.3pt,arc=0.5mm,
  width=0.98\linewidth,left=0.8mm,right=0.8mm,top=0.5mm,bottom=0.5mm]
\scriptsize
\textbf{Root:} \texttt{textiowrapper-tell-reentrant-snapshot-uaf}\\
\textbf{Component:} \texttt{Modules/\_io/textio.c}
\hfill \textbf{Affected:} CPython 3.13--3.15, FT and GIL\\
\textbf{Observed impact:} reclaimed snapshot bytes consumed as live input
\end{tcolorbox}

\noindent\hfill\begin{minipage}{0.96\linewidth}
{\scriptsize\textit{Borrow retained across decoder callbacks:}}\par
\begin{lstlisting}[style=casecode,firstnumber=2795]
PyArg_ParseTuple(self->snapshot, "iO", &flags, &next_input);
...
input = PyBytes_AS_STRING(next_input);
_textiowrapper_decoder_setstate(self, &cookie);
DECODER_DECODE(input, skip_bytes, chars_decoded);
\end{lstlisting}
\end{minipage}\hfill\mbox{}\par

\noindent\textbf{Trigger.}
\texttt{next\_input} is borrowed from \texttt{self->snapshot} across
user-defined decoder methods; the argument-parsing API specifies borrowed
output references~\cite{py-arg-parsing}. A decoder callback re-enters
\texttt{seek(0)}, clearing the snapshot and releasing the bytes. The outer
\texttt{tell()} continues through raw pointers into reclaimed storage. The
object critical section blocks peer threads but not this same-thread callback,
so the dependency remains unprotected.

\noindent\textbf{Validation and security impact.}
The PoC uses re-entrant \texttt{seek} to refill the freed size class with
\texttt{Z}. FT and GIL runs then consume \texttt{Z} through bulk and single-byte
paths, allowing a malicious codec to corrupt stream position or crash the process.

\end{document}